\documentclass[a4paper,fleqn,usenatbib]{mnras}

\usepackage[T1]{fontenc}
\usepackage{ae,aecompl}
\usepackage{subfig}

\usepackage{graphicx}	% Including figure files
\usepackage{amsmath}	% Advanced maths commands
\usepackage{amssymb}	% Extra maths symbols

\title[Major Mergers -- I. Size, shape, and spin]{Major mergers between dark matter haloes -- I. Predictions for size, shape, and spin}

\author[N.E. Drakos et al.]{
	Nicole E. Drakos$^{1,2} $\thanks{E-mail: ndrakos@uwaterloo.ca},
	James E. Taylor$^{1,2}$,
	Anael Berrouet$^{2}$,
	\newauthor
	Aaron S. G. Robotham$^{3}$,
	Chris Power$^{3}$,
	\\
	$^{1}$Waterloo Centre for Astrophysics, University of Waterloo, 200 University Avenue West, Waterloo, ON, N2L\,3G1, Canada \\	
	$^{2}$Department of Physics and Astronomy, University of Waterloo, 200 University Avenue West, Waterloo, ON, N2L\,3G1, Canada \\
	$^{3}$ ICRAR, University of Western Australia, 35 Stirling Highway, Crawley, Western Australia 6009, Australia \\
}

\date{Accepted XXX. Received XXX; in original form XXX}

\pubyear{2019}

\begin{document}
	\label{firstpage}
	\pagerange{\pageref{firstpage}--\pageref{lastpage}}
	\maketitle
	
	\begin{abstract}
	The structural properties of individual dark matter haloes, including shape, spin, concentration, and substructure, are linked to the halo's growth 
	history, but the exact connection between the two is unclear. One open question, in particular, is the effect of major mergers on halo structure. 
	We have performed a large set of simulations of binary equal-mass mergers between isolated haloes with various density profiles, to map out 
	the relationship between the initial conditions and merger parameters and the structure of the final remnant. In this paper we describe our 
	initial set-up and analysis methods, and report on the results for the size, shape, and spin of the merger remnant. The outcomes of mergers 
	are most easily understood in terms of a scaled dimensionless energy parameter $\kappa$ and an angular momentum (or spin) parameter $\lambda$.
	We find that the axis ratio $c/a$ scales roughly linearly with energy $\kappa$ while the axis ratio $c/b$ scales linearly with spin $\lambda$.
	Qualitatively, mergers on radial orbits produce prolate remnants, while mergers on tangential orbits produce oblate remnants. The spin of the 
	remnant can be predicted from angular momentum conservation, while the overall size changes as $\sim \kappa^{-5}$, as expected from 
	self-similar scaling at constant mean density. We discuss potential cosmological applications for these simple patterns. 
	\end{abstract}
	
	%%%%%%%%%%%%%%%%%%%%%%%%%%%%%%%%%%%%%%%%%%%%%%%%%%
	
	%%%%%%%%%%%%%%%%% BODY OF PAPER %%%%%%%%%%%%%%%%%%

\section{Introduction}

There is now compelling evidence for the existence of dark matter over a vast range of scales, from the horizon scales probed by the Cosmic Microwave Background \cite[e.g.][]{planck2018}, to the galactic and sub-galactic scales probed by local dwarf galaxies \cite[see e.g.][and references therein]{mcconnachie2012}. On large scales, dark matter clusters into sheets and filaments; where filaments intersect, they form higher density, roughly spherical structures termed `haloes'. Dark matter haloes are the exclusive sites of galaxy, group and cluster formation. Thus, understanding their structure and evolution is of fundamental importance in cosmology.

Our information about halo structure and evolution comes mainly from $N$-body simulations. These have established that some halo properties, such as the spherically averaged density profile, are approximately universal \citep{navarro1996,navarro1997}, while other properties such as shape, central concentration, or the presence of substructure, vary from system to system. Since observational information from galaxy kinematics \citep[e.g.][]{ouellette2017}, satellite kinematics \citep[e.g.][]{guo2012}, and weak and/or strong gravitational lensing \citep[e.g.][]{umetsu2016} 
is beginning to fix or constrain these individual properties for large samples of haloes, the time is ripe to consider what the structure of individual haloes can teach us.

It is clear that the structure of individual haloes is closely related to their merger history. Shape changes, for instance, have been linked to the parameters of the last major merger \cite[e.g.][]{despali2017}, the remnant being elongated along the merger axis  \citep[e.g.][]{maccio2007,veraciro2011}. Concentration is correlated with the overall age of a halo \citep[e.g.][]{wechsler2002,zhao2009,wong2012}; the density in the central regions reflects the background density of the universe at the time of formation \citep{navarro1996,navarro1997,bullock2001}, or possibly at the end of the rapid, major-merger dominated phase of halo growth \citep[e.g.][]{zhao2003a}. Finally, substructure is formed from the tidally stripped cores of infalling subhaloes, and is thus mainly determined by the recent merger history \citep[e.g.][]{taylor2005b}.

Measuring the shape, concentration, and substructure of haloes may therefore provide an opportunity to learn about individual growth histories, and the connection between growth history 
and the properties of the visible galaxy or galaxies that reside within a halo. Given a quantitative understanding of the connection between history and structure, measurements of structural properties for large, well-defined samples  may also provide new cosmological tests \citep[see e.g.][for a discussion]{taylor2010}. So far, observations of cluster shape \citep[e.g.][]{oguri2010}
and concentration \citep[e.g.][]{sereno2018} have been shown to be consistent with the concordance cosmology established by other tests. Both observational systematics and theoretical predictions need to be refined, however,  before these methods can be used to improve our knowledge of the cosmological parameters.

Halo growth occurs through accretion of material from the surrounding density field, both smoothly and in a series of violent, stochastic mergers. To understand the smooth part of the process requires cosmological simulations, in order to capture the statistics of the density and velocity fields around the peaks where haloes form. There have been extensive theoretical studies of halo structure in this cosmological context, though many focus on mean trends, rather than individual cases \cite[e.g.][]{navarro1997, bullock2001, zhao2003a, butsky2016, klypin2016, despali2017}.
Mergers complicate the picture, however; sufficiently so that they have often been studied in simpler, idealized simulations with controlled initial conditions \citep[ICs; e.g.][]{fulton2001, boylankolchin2004, moore2004, mcmillan2007, vass2009, ogiya2016}.

Even when studying major mergers using isolated simulations, there are still many degrees of freedom (halo profile, mass ratio, shape, orbit) that can obscure which essential parameters determine the properties of the final remnant. Therefore, in this paper we will start by considering the simplest case: an equal-mass merger between two isolated, identical spherical systems, 
given one of various realistic density profiles, and placed on a variety of initial orbits. This work is the first in a series; in this paper we will introduce our major merger simulations and examine how the ICs determine the shape and spin of the final remnant. In subsequent papers, we will examine how other properties, such as concentration and the detailed form of the density profile, depend on the ICs of the merger. 

The outline of the paper is as follows. In Section~\ref{sec:ICs} we describe the initial halo models, and verify their stability in isolation. In Section~\ref{sec:SetUp} we describe the set-up and analysis of the merger simulations. In Section~\ref{sec:Res} we present our main results on shape and spin. Finally, in Section~\ref{sec:Discuss} we discuss our conclusions, the limitations of this study, and future work. 

 %%%%%%%%%%%%%%%%%%%%%%%%%%%%%%%%%%%%%%%%%%%%%%%%%%%%%%%%%%%%%%%%%%%	 

\section{Halo Models}\label{sec:ICs}

Each of our simulations follows the merger of two identical haloes. To investigate the effect of the halo model on the shape and spin of the remnant, we consider several different initial models, as described below.

\subsection{Initial conditions}

We consider initial halo models with NFW \citep{navarro1996,navarro1997} and Einasto \citep{einasto1965} profiles. The ICs were created using the code \textsc{icicle} \citep{drakos2017}. Since the mass of an NFW profile diverges at large radii, the ICs need to be truncated in order to be realized with a finite number of particles. One common approach is to use an exponentially truncated NFW profile (hereafter denoted ``NFWX''), which is NFW within the virial radius, $r_{\rm vir}$, and then decays exponentially outside the virial radius \citep{springel1999}. An additional parameter $r_{\rm decay}$ sets how fast the decay occurs with radius. An alternative approach to truncating an NFW profile is to generate the part of the profile interior to some tidal radius, $r_t$, and then iteratively remove any particles that are unbound, given the escape speed of the truncated system. As the tidal radius approaches infinity, systems generated in this way are equivalent to an infinitely extended NFW profile. It has been shown that the profiles resulting from this second approach are stable, and resemble tidally stripped NFW profiles \citep{drakos2017}, and thus we denote them ``NFWT'' profiles.

Overall, we considered six different initial profiles. Four of these profiles are NFW, but truncated in different ways (two are NFWT and two are NFWX profiles).  Two are Einasto profiles, with $\alpha_E$ values representative of the range found in simulations \citep{gao2008}: to explore how the inner slope affects the simulations we use an Einasto profile with a low $\alpha_E$ value of $0.15$, and also a profile with a high $\alpha_E$ value of 0.3. 

\begin{figure}
	\includegraphics[width=\columnwidth,scale = 0.7,trim={0cm 0cm 6cm 0},clip]{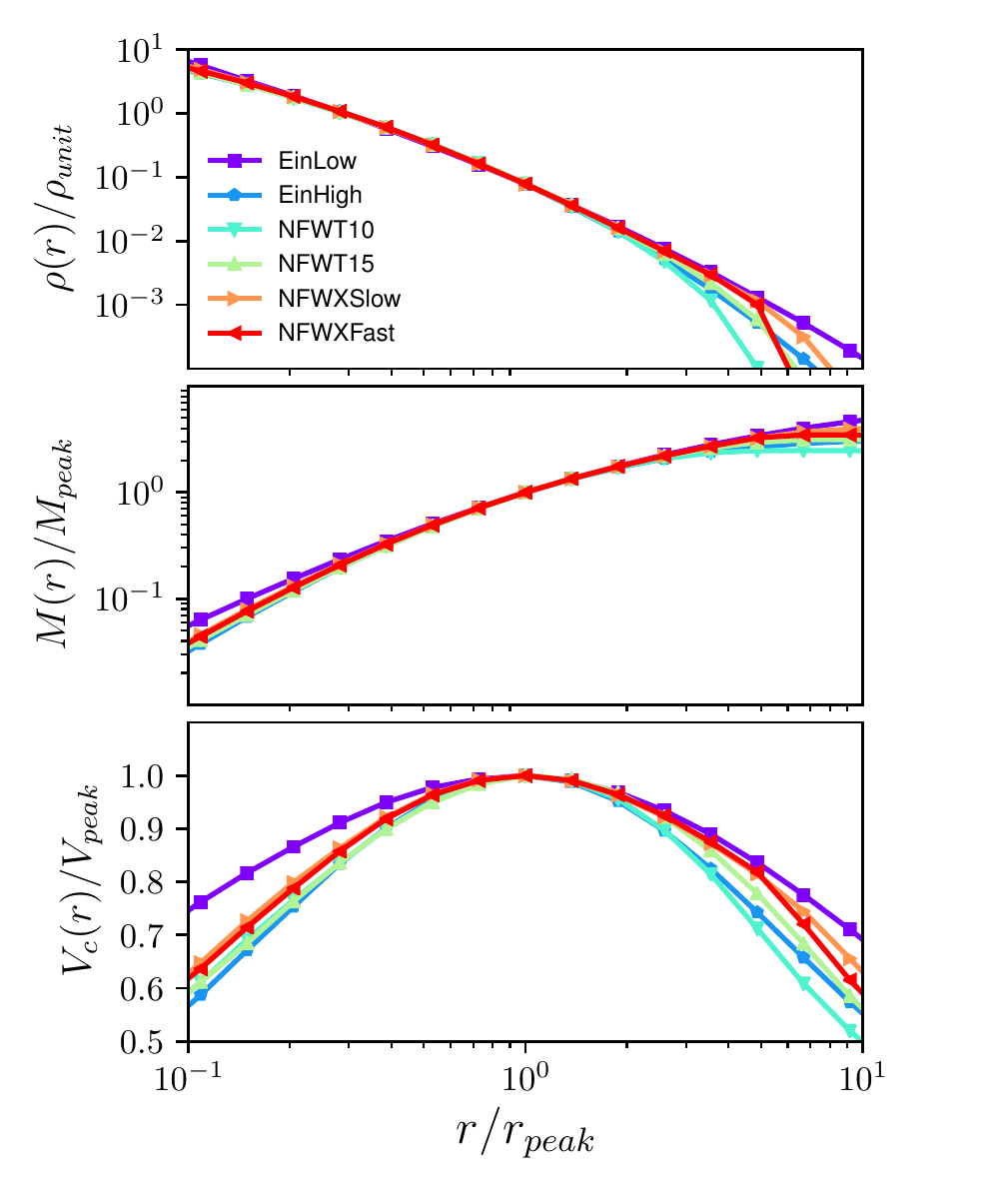}.
	\caption{Comparison of the density (top), enclosed mass (middle), and circular velocity (bottom) profiles of the initial models. }
	\label{fig:Profiles}
\end{figure}

The simulation units were chosen so that the gravitational constant, $G$, the peak circular velocity, $v_{\rm peak}$, and the radius at which the circular velocity peaks, $r_{\rm peak}$, are all unity. Setting $G=M_{\rm peak}=r_{\rm peak}=1$ produces a time unit $t_{\rm unit} = \sqrt{r_{\rm peak}^3/GM_{\rm peak}}$, a density unit $\rho_{\rm unit}=M_{\rm peak}/r_{\rm peak}^3$ and an energy unit $E_{\rm unit} = GM_{\rm peak}^2/r_{\rm peak}$. All the haloes were constructed initially using $5 \times 10^5$ particles; after removing unbound particles, the resulting NFWT profiles then have fewer particles. The IC profiles are compared in Fig.~\ref{fig:Profiles}, and the IC parameters and properties are summarized in Table~\ref{tab:ICs}.

%%%%%%FIGURE 1%%%%%%

\begin{table*} 
	\caption{\label{tab:ICs}Summary of the parameters used for the ICs. The profiles are shown in Fig.~\ref{fig:Profiles}. The columns list (1) the name of the ICs, (2) the number of particles $N$, (3) the parameters used to construct the ICs, (4) the relaxation time evaluated at the peak radius, $t_{relax}(r_{\rm peak})$,
		and (5) the total internal energy of the halo, $E_0$.}
	\begin{tabular}{c c c c c c}
		\hline
		Initial conditions name & N &Parameters & $t_{relax}(r_{\rm peak}) / t_{\rm unit}$& $E_0/E_{\rm unit}$\\ 
		\hline
		EinLow & $5 \times 10^5$& $\alpha_E=0.15$ & 610 &  -2.2\\
		EinHigh& $5 \times 10^5$ & $\alpha_E=0.3$ & 1300 &  -1.2\\
		NFWT10& $\sim 3.2 \times 10^5$ & $r_{cut} = 10$ & 1100 & -1.0\\	
		NFWT15& $ \sim 3.5  \times 10^5$ & $r_{cut} =15$ & 1000 &  -1.3\\	
		NFWXSlow& $5 \times 10^5$ & $r_{\rm vir}=10$, $r_{\rm decay} = 2 \, r_s$&1100 &  -1.6\\	
		NFWXFast& $5 \times 10^5$ &  $r_{\rm vir}=10$, $r_{\rm decay} = 0.2 \, r_s$& 1200 &  -1.5\\			
		\hline
	\end{tabular}	
\end{table*}

\subsection{Internal energies}

The internal energy of each halo, $E_0$, was calculated as follows:
\begin{equation}
E_0 = P_0 + K_0  \,\,\, ,
\end{equation} 
where $P_0$ and $K_0$ are the potential and kinetic energy of the halo, respectively. These are most generally expressed as:
\begin{equation}
\begin{aligned}
K_0 &= \sum_{i=1}^N m v_i^2 \\
P_0 &= -\dfrac{1}{2} \sum_{i,j =1}^N \dfrac{Gm^2 }{r_{ij}} \,\,\, ,
\end{aligned}
\end{equation}
where $m$ is the mass of each particle and $r_{ij}$ is the distance between particles $i$ and $j$. Since the ICs are spherically symmetric, however, at least to within the discreteness noise of the individual particles, we can treat the mass of each particle $i$ as being distributed over a shell of radius $r_i$, and write
\begin{equation} \label{eq:P_0}
P_0 \approx -\dfrac{Gm^2}{2}\sum_{i=1}^N\left( \dfrac{N(<r_i)}{r_i} + \sum_{j=1,\\ r_j>r_i}^N \dfrac{1}{r_j} \right) \, ,
\end{equation}
where the two terms in parentheses give the contributions interior to and exterior to the position of each particle $i$, respectively.
The internal energy for each of the halo models is listed in Table~\ref{tab:ICs}.

\subsection{IC stability}

\begin{figure*}
	\includegraphics[scale = 0.9,clip]{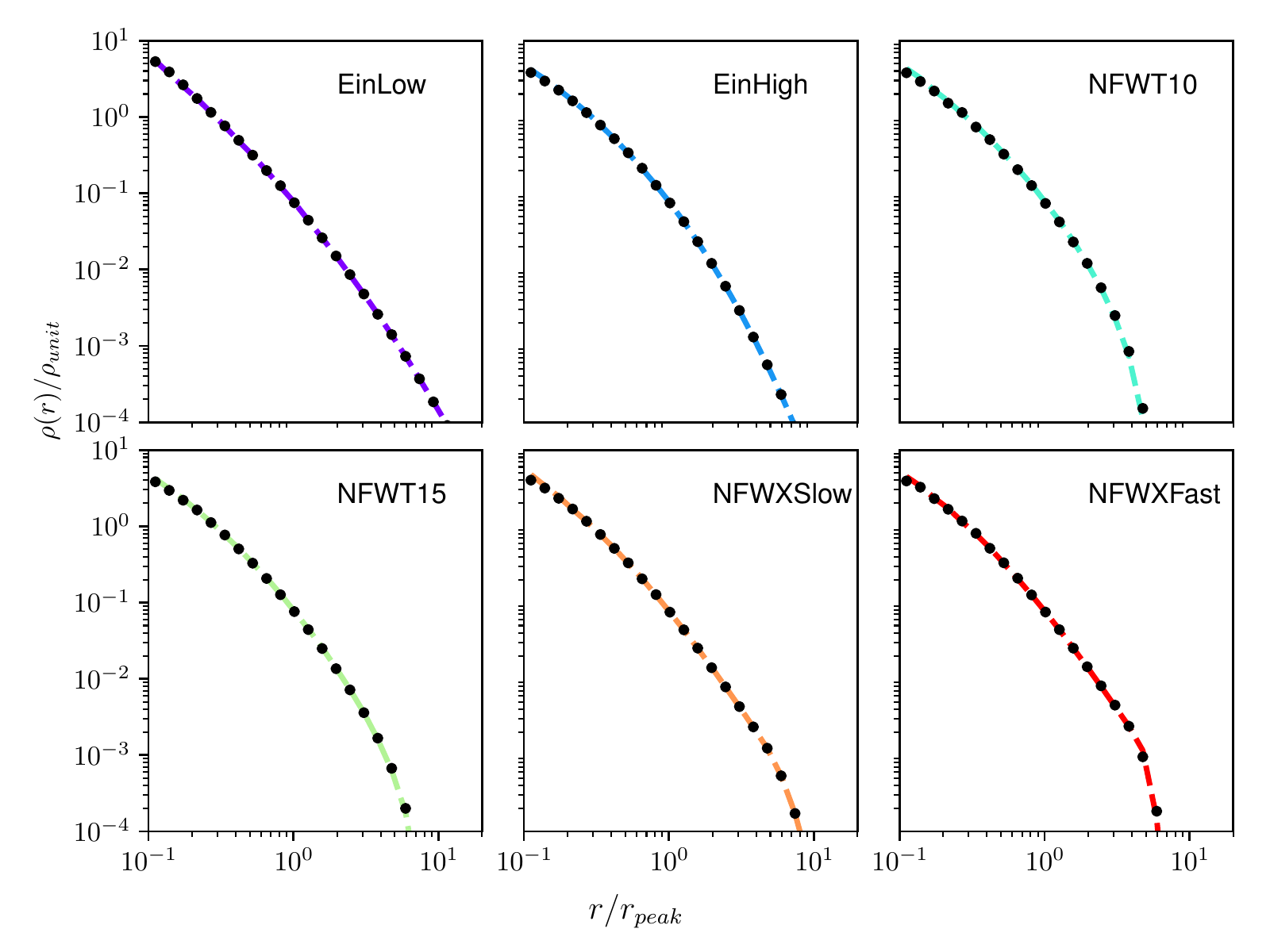}.
	\caption{Stability of the ICs. The dashed line shows the initial profile, while black circles show the profile at the time $t = 300\ t_{\rm unit}$.}
	\label{fig:IC_Stability}
\end{figure*}

To verify the stability of the ICs, they were evolved  in isolation using \textsc{gadget-2} \citep{gadget2}, with a softening length of $\epsilon = 0.02\, r_{\rm peak}$. The stability of the ICs will be limited by relaxation due to the limited number of particles. The characteristic relaxation time for each profile was calculated as follows:
\begin{equation}
t_{\rm rel}(r) = 0.1 \dfrac{\sqrt{N(<r)}}{\ln N(<r)}\sqrt{\dfrac{r^3}{GM(<r)}}\,\,\, ,
\end{equation}
as in \cite{binney}, and is included in Table~\ref{tab:ICs}. Fig.~\ref{fig:IC_Stability} illustrates the stability of the host haloes over a time-scale of $t = 300\ t_{\rm unit}$. We also tested the sensitivity of these results to time step size and the error in the force calculations by increasing and decreasing the \textsc{Gadget-2} parameters \textsc{ErrTolIntAccurac} and  \textsc{ErrTolForceAcc} by a factor of 3, but found that this did not make a noticeable difference in the stability of the ICs.

%%%%%%FIGURE 2%%%%%%

\section{Merger Simulations}\label{sec:SetUp}

The merger simulations were run in \textsc{Gadget-2} using $N=5 \times 10^5$ particles per halo and a softening length of $\epsilon= 0.02\, r_{\rm peak}$. The centre of the remnant halo was found by calculating the centre of mass within increasingly smaller spheres; as in \cite{moore2004}, we found that this was roughly equivalent to tracking the most bound particle.

%%%%%%%%%%%%%%%%%%%%%%%%%%%%%%%%%%%%%%%%%%%%%%%%%%%%%%%%%%%%%%%%%%%
\subsection{Orbital parameters}

For each of the six models, we performed 30 equal-mass binary merger simulations, for a total of 180 simulations. Simulations were analyzed in the rest frame of the first halo, the second halo being given an initial position and velocity in this frame. We considered three different radial separations, $r_{\rm sep} = 2, 5$, or $10 \, r_{\rm peak}$. The initial velocity was either purely tangential or purely radial, with magnitude $v_0 =  0.1,\, 0.2,\, 0.6,\, 0.8 \, $ or $1.2 \, v_{\rm esc}$, where $v_{\rm esc}$ is the escape velocity of a point mass located at $r_{\rm sep}$. 

Each orbit can be described by its energy and angular momentum. The orbital energy, $E_{\rm orb}$, was calculated as the total energy of the system minus the internal energy of the individual haloes,  i.e.,
\begin{equation}
E_{\rm orb} =  P + K  - 2E_0 \,\,\, ,
\end{equation} 
where the internal energy of each halo, $E_0$, was calculated as described in Section~\ref{sec:ICs}, and the total potential and kinetic energies of the system were calculated as described in \cite{bett2010}:
\begin{equation} \label{eq:KandP}
\begin{aligned} 
K &= \dfrac{1}{2} \sum_{i=1}^N m \mathbf{v_i}^2 \,\,\,, \\
P &= \left(\dfrac{N^2-N}{N_{\rm sel}^2-N_{\rm sel}}\right) \left(\dfrac{-Gm^2}{\epsilon}\right) \sum_{i=1}^{N_{\rm sel} -1} \sum_{j=i+1}^{N_{\rm sel}} -W(r_{ij}/\epsilon) \, .
\end{aligned}
\end{equation}
Here $N_{sel}$ is the number of randomly selected particles, used to approximate the entire distribution; after experimentation we found that 5000 particles from each halo were sufficient to calculate $P$ accurately. $W$ is the smoothing kernel used for force calculations in \textsc{Gadget-2}  \citep{springel2001},
\begin{equation} 
W(x) =
\begin{cases}
\dfrac{16}{3}x^2 - \frac{48}{5}x^4 + \frac{32}{5} x^5 -  \frac{14}{5}, & 0 \leq x \leq  \frac{1}{2}\\
\frac{1}{15x} + \frac{32}{3}x^2 - 16 x^2 +\frac{48}{5}x^4\\ -\frac{32}{15}x^5 - \frac{16}{5}, 
& \frac{1}{2} \leq x \leq  1\\
-\frac{1}{x}, &  x \geq 1 \,\,\, .
\end{cases}
\end{equation}

\begin{figure}
	\includegraphics[width = \columnwidth,trim={3cm 4cm 3cm 0},clip=true]{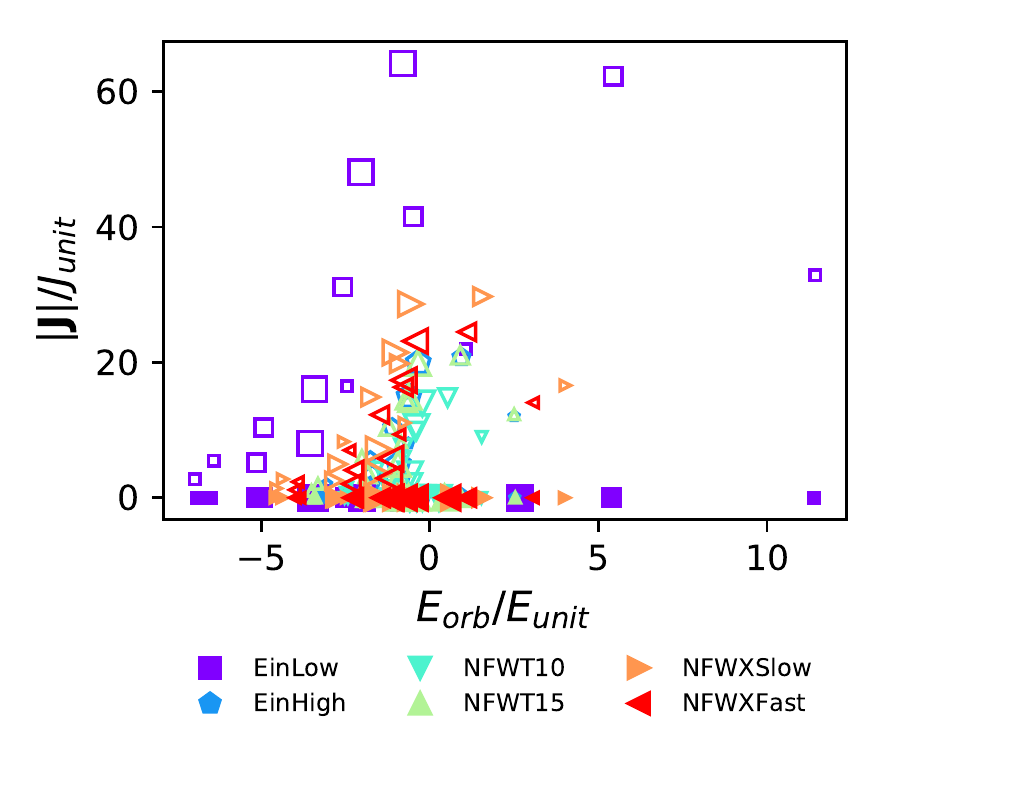}.
	\caption{The range of orbital energies $E_{\rm orb}$ and angular momenta $J$ used in the simulations. Colours and symbols are as in Fig.~\ref{fig:Profiles}. Open points indicate tangential initial velocities, and filled points denote radial initial velocities. The size of the symbols indicates the initial radial separation, $r_{\rm sep}$.
	}	
	\label{fig:OrbitalParameters}
\end{figure}

Finally, the orbital angular momentum was calculated as follows:
\begin{equation} \label{eq:L}
\mathbf{J} = \sum_i m \mathbf{r}_i \times \mathbf{v}_i \,\,\, ,
\end{equation} 
which was found to be equivalent to $\mathbf{J} = M \mathbf{r}_{\rm sep} \times \mathbf{v_0}$ to within $0.5$ per cent. The energy and angular momenta used in our simulations are shown in Fig.~\ref{fig:OrbitalParameters}.

%%%%%%FIGURE 3%%%%%%

In cosmological simulations, orbital energy is often expressed in units of the energy of a circular orbit at the virial radius, and angular momentum is expressed in terms of the circularity, $\epsilon$, defined as the angular momentum divided by the angular momentum of a circular orbit with the same energy, $\epsilon=J/J_C(E)$ \citep{lacey1993}. Neither of these quantities is well defined in this context, however, since the virial radius does not have a clear definition for isolated, non-cosmological simulations, and the definition of circularity requires that the orbit be bound. 

Overall, our simulations cover a wider range of energy and angular momentum than would be expected in a cosmological scenario. Typically, subhaloes merging into galaxy- or cluster-sized haloes have a broad distribution of orbital circularity with a mean of $\epsilon \approx 0.5$ (although primordial haloes may merge on more radial orbits; \cite{ogiya2016}) while the energies of cosmological orbits are typically close to that of a circular orbit at the virial radius  \citep[e.g.][]{khochfar2006,wetzel2011,jiang2015}. In contrast, the orbits in our simulations are chosen to sample the full range of physical possibilities, in order to determine how orbital parameters affect the outcome of a merger generally. Thus we include completely radial and completely tangential orbits, and also a broad range of possible energies, from almost unbound to extremely tightly bound.

\subsection{Merger time-scale}

\begin{figure}
	\includegraphics[width = \columnwidth,trim={0cm 5cm 6cm 0},clip]{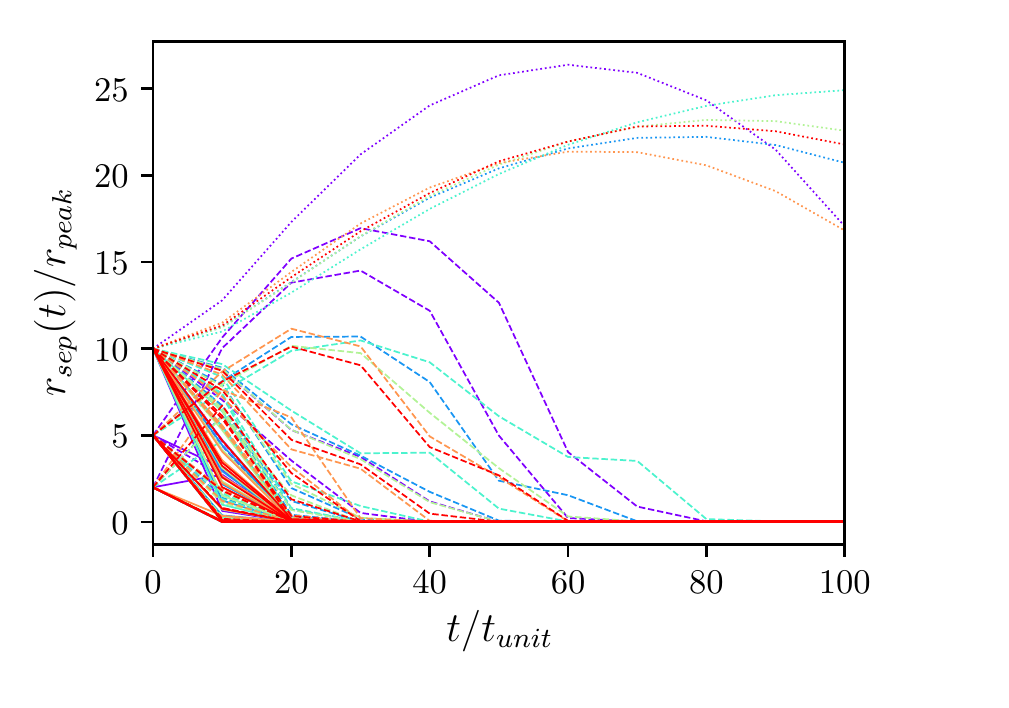}.
	\caption{Radial separation between the halo centres, as a function of time. The six cases that did not merge by $t = 100\,t_{\rm unit}$ (dotted lines) 
	were excluded from subsequent analysis. Colours indicate the initial profile model, as in Fig.~\ref{fig:Profiles}. Dashed lines indicate tangential initial velocities, and solid lines denote radial initial velocities.}
	\label{fig:Frame_vs_Time}
\end{figure}

Fig.~\ref{fig:Frame_vs_Time} shows the radial separation between the merging haloes as a function of time. At the highest relative velocity, haloes on tangential orbits had not merged by the time $t=100\, t_{\rm unit}$, and thus were excluded from this study, such that only 174 simulations are analyzed. For the majority of the simulations, the haloes merged very quickly (i.e.~in less than one orbit). This is broadly consistent with predicted orbital decay times due to dynamical friction \citep{colpi1999}:
\begin{equation}
\tau_{\rm DF} = 1.2 \dfrac{J_{\rm circ} r_{\rm circ}}{(GM_{\rm sat}/e)\log(M_{\rm halo}/M_{\rm sat})} \epsilon^{0.4} \,\,\, ,
\end{equation}
where $J_{\rm circ}$ is the angular momentum of a circular orbit with the same energy, $\epsilon$ is the circularity (i.e.~the angular momentum divided by the angular momentum of a circular orbit of the same energy), and $r_{\rm circ}$  is the radius of a circular orbit with the same energy. $M_{\rm sat}$ and $M_{\rm halo}$ are the masses of the satellite halo and the main system, respectively. Although this equation cannot be applied directly to the majority of our simulations (since many of our orbits are unbound, and because this estimate is not applicable to completely radial orbits), it predicts decay times of less than an 
orbital period as the mass ratio approaches unity.

%%%%%%FIGURE 4%%%%%%

\begin{figure}
	\includegraphics[width = \columnwidth,trim={0 0cm 6cm 0},clip=true]{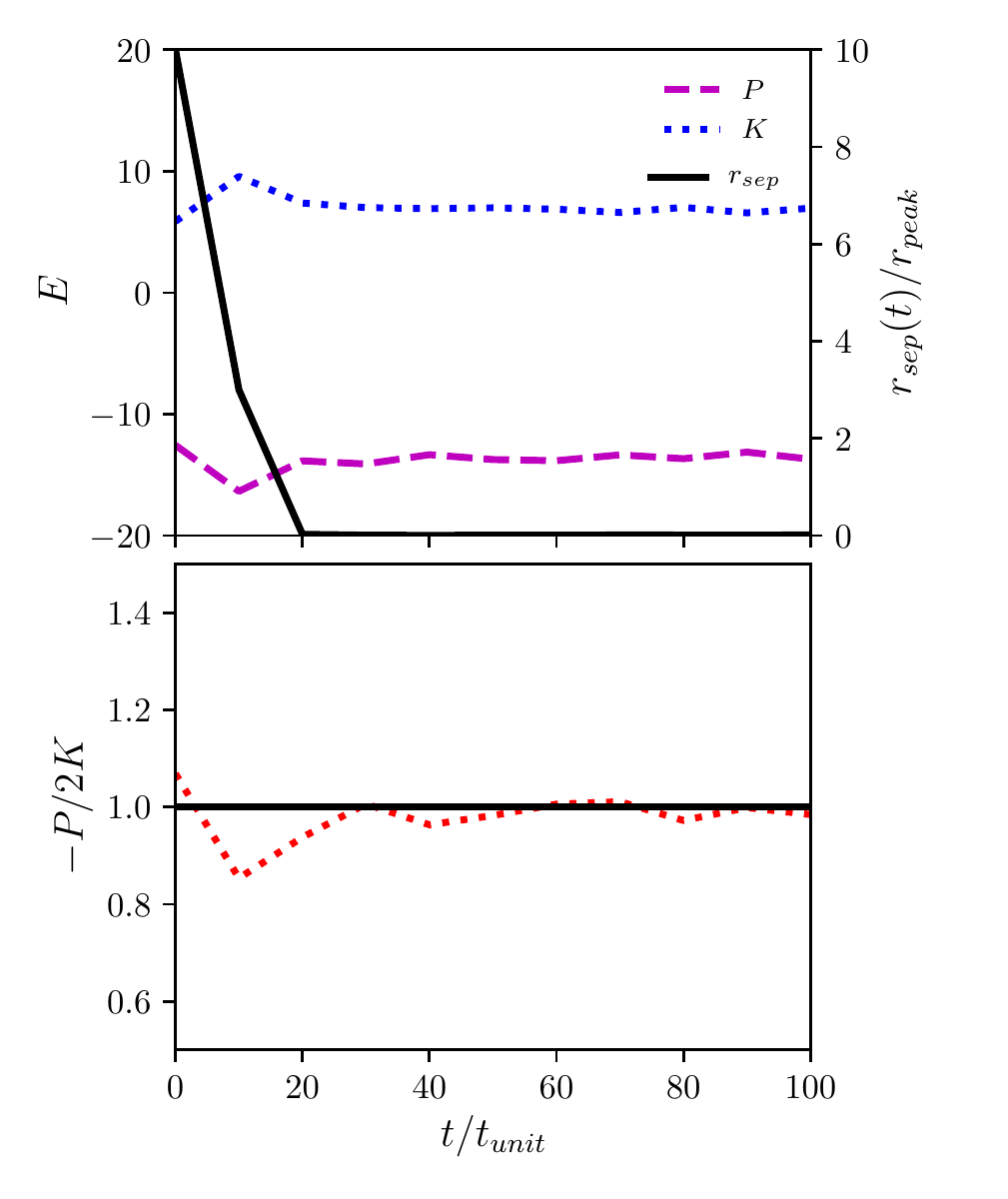}.
	\caption{Evolution towards equilibrium with time. The top panel shows the radial separation between the halo centres (solid black line), as well as the potential and kinetic energy of the entire system. The bottom panel shows the virial ratio $-P/2K$ (dotted red line), which should be 1 for a virialized system (solid black line).}
	\label{fig:Energy_vs_Time}
\end{figure}

We also determined how long it took for the remnant to reach virial equilibrium, after the merger. At every time step the potential and kinetic energy, $P$ and $K$, were calculated in the frame of the first halo, using equation~\eqref{eq:KandP}. Fig.~\ref{fig:Energy_vs_Time} shows an example of the time evolution of these energies, the virial ratio, and the separation between the halo centres, in a simulation with EinLow ICs, with orbital parameters $r_{\rm sep} = 10\, r_{\rm peak}$ and a radial velocity of $v_0= 0.8\,v_{\rm esc}$. For this simulation, the haloes merge by $t = 20 \,t_{\rm unit}$, which is less than one period of the initial orbit. The potential and kinetic energy remain constant after the halo has merged, with a virial ratio of $-P/2K \approx 1$ to good approximation. Although we only demonstrate that virial equilibrium is reached when $r_{\rm sep} \approx 0$ for one case, this result holds for all the simulations in this work.

%%%%%%FIGURE 5%%%%%%

\subsection{Shape measurement} \label{sec:shape_measure}

A halo is, in general, triaxial, with  principal axes $a>b>c$. Prolate haloes will have one long and two short axes, while oblate haloes will have two long and one short axes. Following \cite{moore2004}, we calculated principal axis ratios $s = b/a$ and $q=c/a$ using the iterative method described in \cite{dubinski1991}. Beginning with $a=b=c=1$, the eigenvalues $w_1 = a^2$, $w_2 = b^2$, and $w_3 = c^2$ were found from the dimensionless inertia tensor $I_{ij} = \sum x_i x_j /d^2$, where $d = x_i^2 + (x_j/s)^2 + (x_k/q)^2$ is the ellipsoidal coordinate. The coordinates of each particle were then rotated using the eigenvectors of the new principal axis, and the principal axis ratios were recalculated. This process was repeated until convergence, which was defined as when $s$ and $q$ both had a relative change of less than $10^{-5}$.

\begin{figure*}
	\includegraphics[scale = 0.9, trim={0 2cm 0 2cm},clip=true]{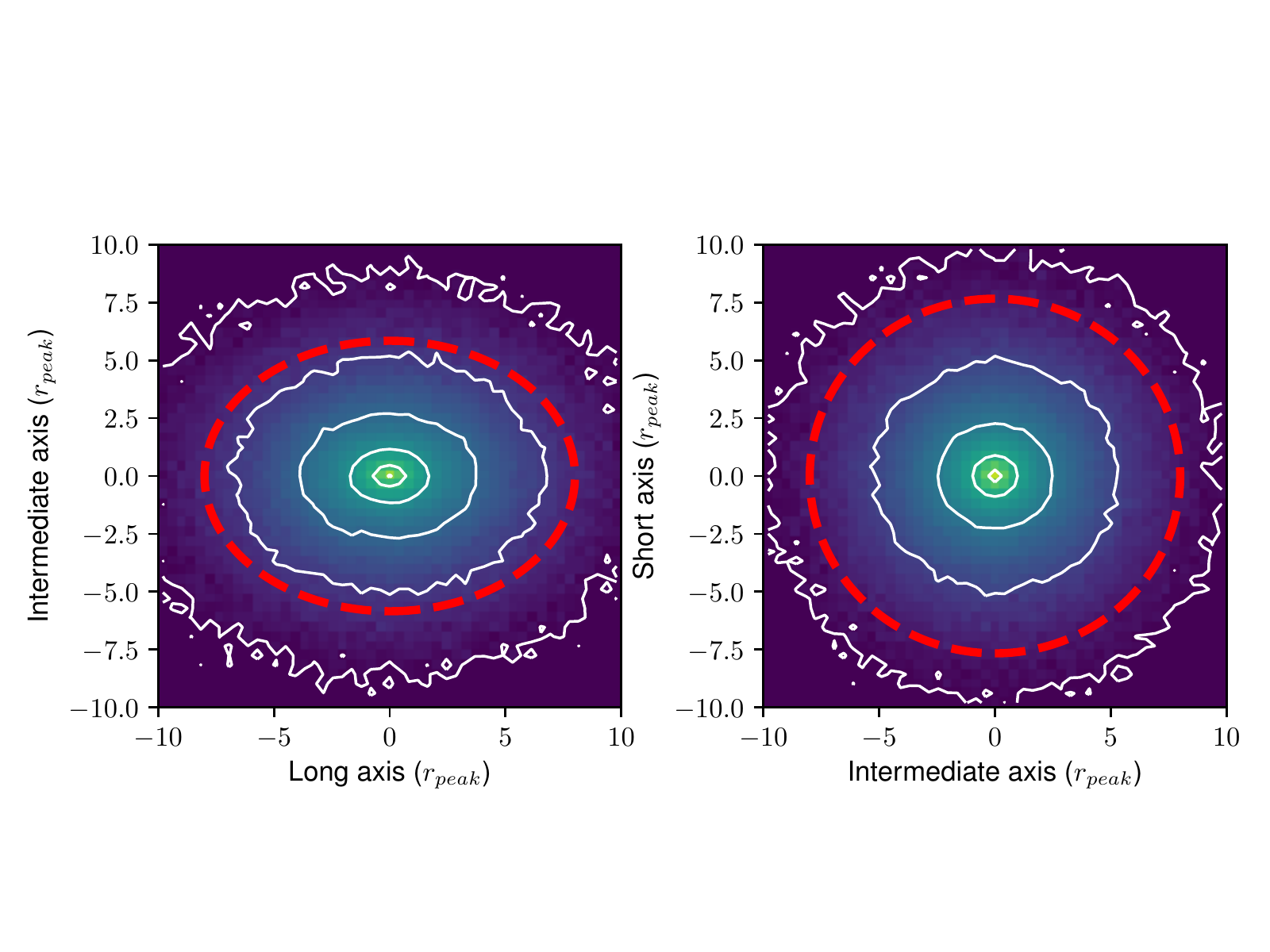}.
	\caption{Projected isodensity contours (white lines) of the halo remnant. The measured shape ratio is shown as by the red dashed line.}
	\label{fig:Bin2D}
\end{figure*}

In Fig.~\ref{fig:Bin2D} we compare the shape measurement to the isodensity contours of the remnant halo for a radial merger between two haloes with EinLow profiles. The haloes were initially separated by $r_{\rm sep} = 10 \,r_{\rm peak}$, and the second halo was given an initial velocity of $v_0 \approx 0.9\, v_{\rm peak} $ (this corresponds to 80 per cent of the escape speed of a point mass located at $r_{\rm sep}$). The shape measurement agrees well with the isodensity contours. We note that since our shapes are measured using all the halo particles, we will not catch radial variations in halo shape, but at least in this case, radial variations do not appear to be significant.

%%%%%%FIGURE 6%%%%%%

We investigated the effect of numerical resolution on our shape measurements by considering haloes with various numbers of particles, ranging from $N=5 \times 10^3$ to $N=5 \times 10^5$. The original softening length of  $\epsilon= 0.02\, r_{\rm peak}$ for the $N=5\times 10^5$ profile was scaled as $N^{-1/3}$. As above, we considered the case of a merger between two EinLow profiles, since this profile has the shortest relaxation time. 

\begin{figure}
	\includegraphics[width = \columnwidth,trim={3cm 0cm 4cm 0},clip=true]{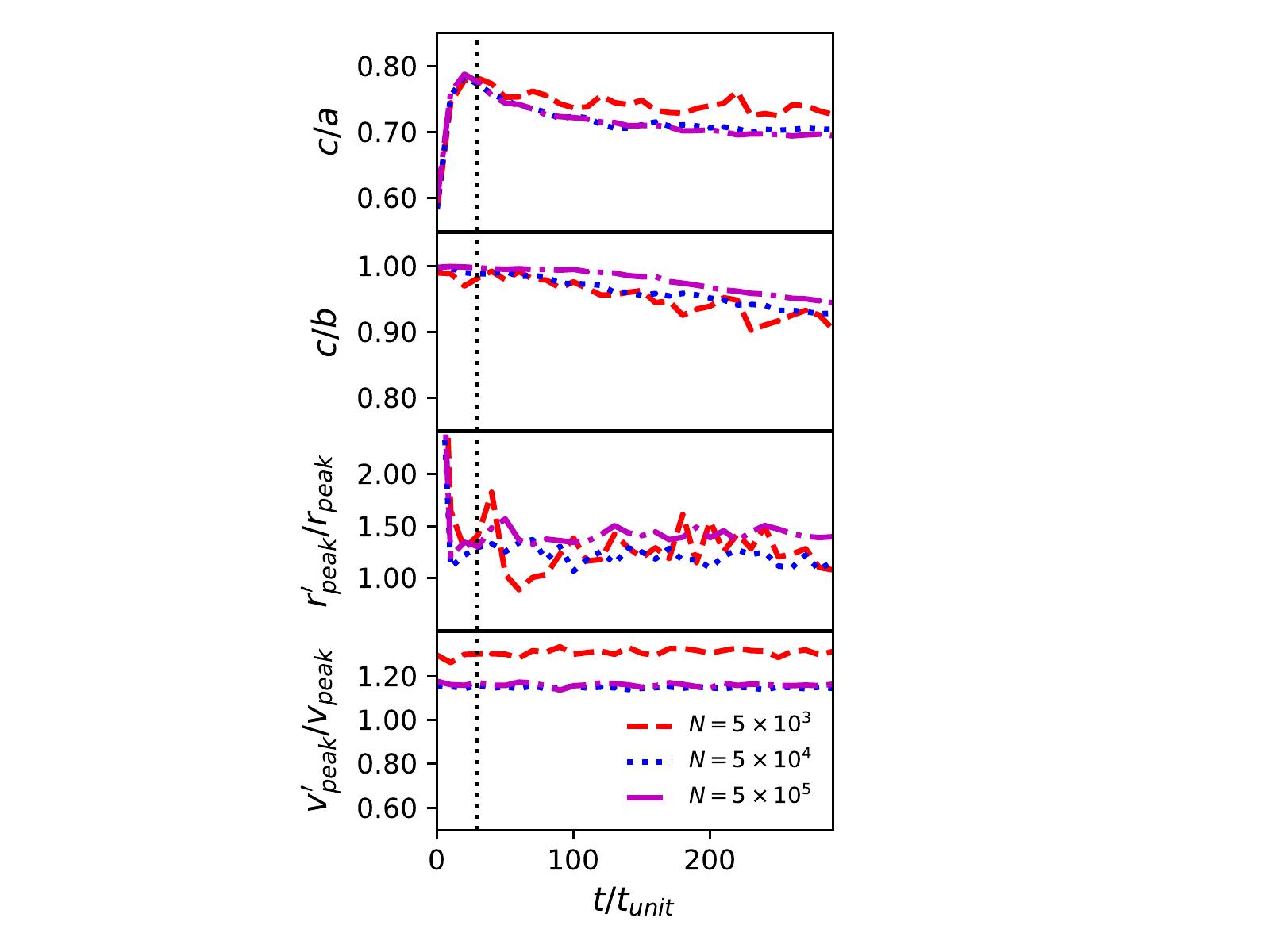}.
	\caption{The axis ratios $c/a$ and $c/b$ and the structural parameters $r'_{\rm peak}$ and $v'_{\rm peak}$ relative to their original values $r_{\rm peak}$ and $v_{\rm peak}$, as a function of time, for different resolutions. The vertical dotted lines show the time by which the two haloes had completely merged.}
	\label{fig:TimePlot}
\end{figure}

We compared shape parameters $c/a$ and $c/b$, as well as the values of $r_{\rm peak}$ and $v_{\rm peak}$. Fig.~\ref{fig:TimePlot} shows 
how these properties change as a function of time at each resolution. While there are significant fluctuations in the values over time at 
low resolution, at the resolution of our main set of simulations, the structural parameters are stable between $t=100$ and $300\, t_{\rm unit}$. 

\begin{figure}
	\includegraphics[width = \columnwidth,trim={3cm 0cm 4cm 0},clip]{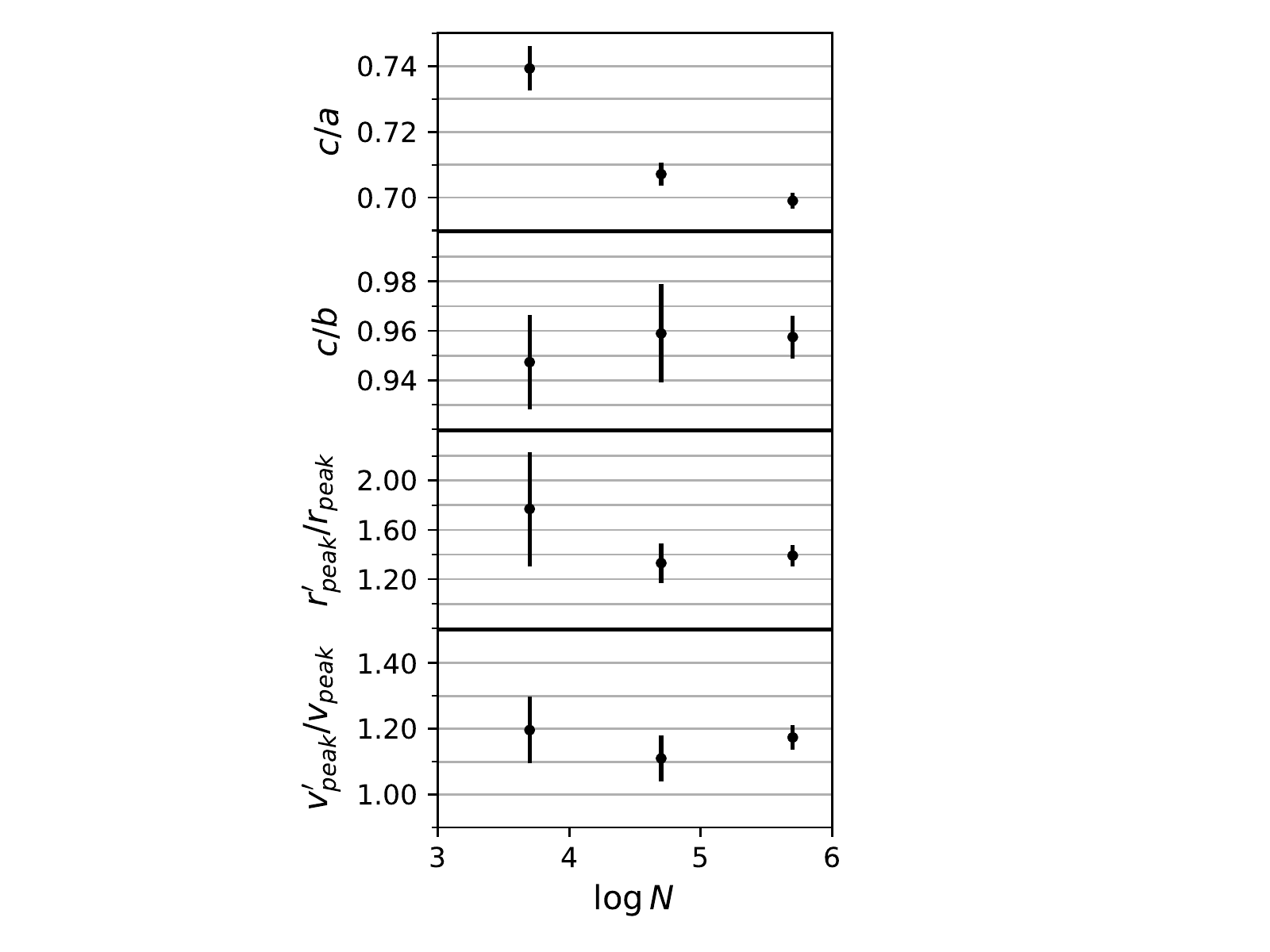}.
	\caption{The axis ratios $c/a$ and $c/b$ and the structural parameters $r'_{\rm peak}$ and $v'_{\rm peak}$ relative to their original values $r_{\rm peak}$ 
	and $v_{\rm peak}$, as a function of resolution. Each point is averaged over five realizations, while error bars show the r.m.s..}
	\label{fig:ErrorBars}
\end{figure}

The net effect of resolution on the final measurements (averaged over 10 snapshots between $t = 200\, t_{\rm unit}$ and $300\, t_{\rm unit}$) is shown in Fig.~\ref{fig:ErrorBars}. For each point, five different realizations were simulated, using different random seeds to generate the ICs. Low resolutions tend to predict more circular haloes, but at the resolution of $5\times10^5$ particles per halo, the shape parameters are determined to within 1 per cent or better. The maximum of the circular velocity, $v_{\rm peak}$ has similar accuracy, while $r_{\rm peak}$ is more sensitive to relaxation, and has an uncertainty of 6 per cent. We note that there is a transient period before the haloes merge in which $c/a$ is very small; this is because the algorithm we use for calculating shapes takes into account the particles of both (un-merged) haloes.

%%%%%%FIGURE 7%%%%%%

%%%%%%FIGURE 8%%%%%%

We conclude that a resolution of $5 \times 10^5$ particles per halo is sufficient to study the properties of the merger remnant. At lower resolution, numerical relaxation can artificially increase the location of $r_{\rm peak}$. We found that this effect could be somewhat alleviated by decreasing 
the time-stepping parameter (\textsc{ErrTolIntAccuracy} in \textsc{Gadget-2}), at the cost of much slower run times. For the simulations shown 
in this paper, the value \textsc{ErrTolIntAccuracy}$=0.02$ was used.

\subsection{Halo rotation}

Finally, we check to see whether the halo remnants are in solid-body rotation. In the previous section, we showed that the axis ratios stayed approximately constant with time, after an initial transient period. This result holds for all the simulations performed in this work. Since the merger 
remnants from tangential encounters should rotate due to conservation of the initial angular momentum, we are also interested in whether they
rotate differentially, or as a solid body. We considered cases in which the remnant is prolate ($b/a<0.8$), such that the major axis has a well-defined direction, and measured rotation by tracking changes in the orientation of the major axis.

\begin{figure}
	\includegraphics[width = \columnwidth,trim={3cm 0cm 3cm 0},clip]{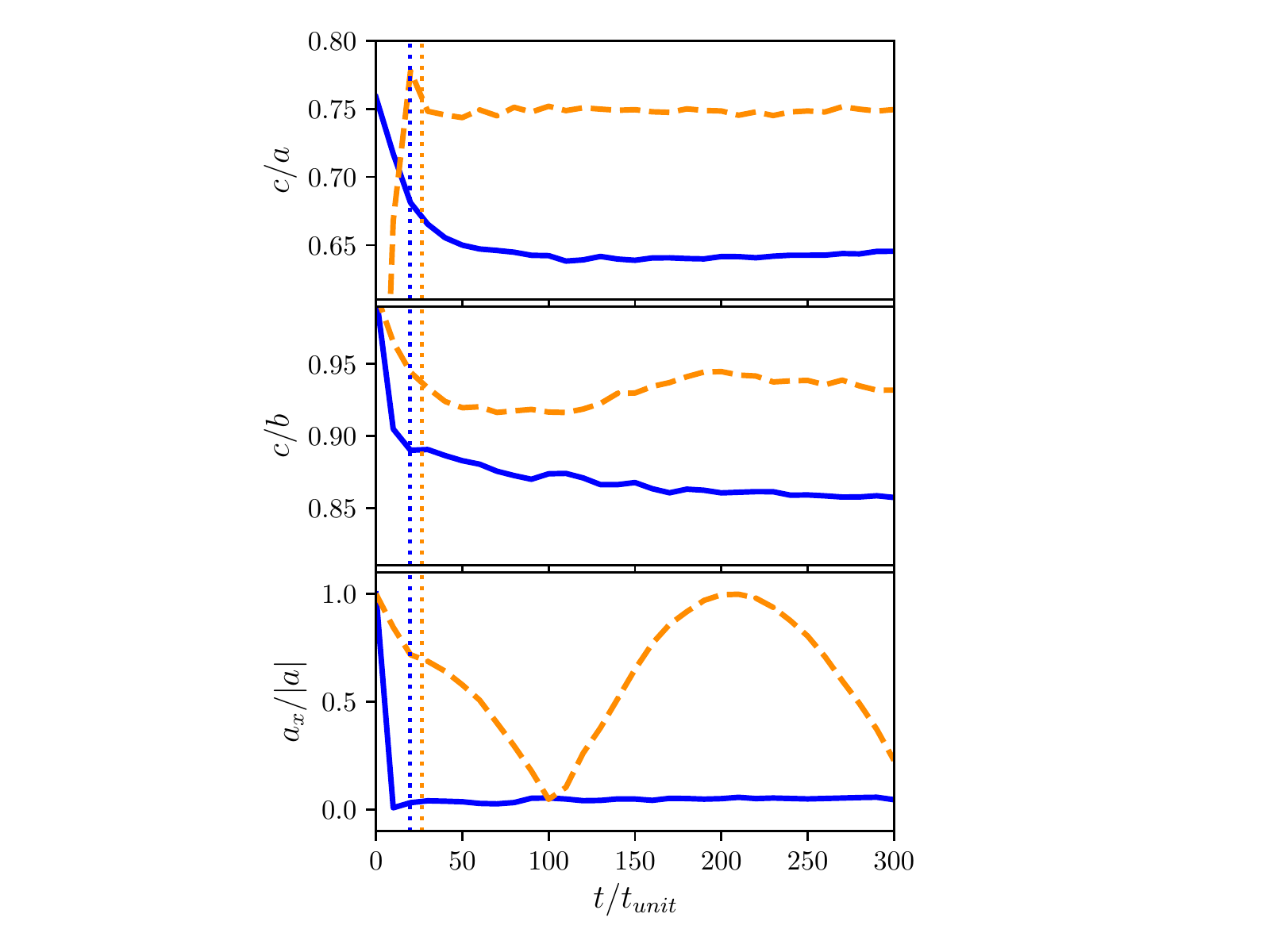}.
	\caption{The time evolution in the shape ratios $c/a$ and $c/b$ (top and middle panels), as well as the $x$-component of the normalized major axis (bottom panel). The two examples are mergers between two EinHigh profiles (orange dashed lines) and two EinLow profiles (blue solid lines), chosen because they have comparable energies and angular momenta. The vertical dotted lines show the time by which the two haloes had completely merged.}
	\label{fig:Rotation_vs_Time}
\end{figure}

In Fig.~\ref{fig:Rotation_vs_Time} we show two sample cases, chosen such that the mergers have similar energies and angular momenta. The first is a merger between two EinHigh profiles with orbital parameters $r_{\rm sep}=5\, r_{\rm peak}$ and $v_0 =0.2\, v_{\rm esc}$. The second case consists of two EinLow profiles with orbital parameters $r_{\rm sep}=2\, r_{\rm peak}$ and $v_0 =0.8\, v_{\rm esc}$. The top two panels show the axis ratios, while the bottom panel shows 
the normalized $x$-component of the major principal axis, $a_x/ |\mathbf{a}|$, as a measure of orientation.  As in Figure~\ref{fig:TimePlot}, $c/a$ is very small in the transient period before the haloes merge since our algorithm calculates shape based on all the particles from both haloes.

In the first case, the halo appears to be rotating as a solid body, and there may be some slight change in the shape ratio $c/b$, though the change is within the uncertainty of the shape measurements. In the second case, there is no clear rotation. The main difference between these two cases is that the EinLow profile is very extended, with a lot of mass at large radii; it is possible that in this case the envelope is decoupled from the core, explaining the irregular variation in the orientation of the major principal axis. We conclude that the rotation of the remnants can be complicated, including both differential and solid-body rotation. In either of the two cases, however, the long-term shape of the remnants is well defined. We will proceed to study how this shape depends on halo profiles and orbital parameters.

%%%%%%FIGURE 9%%%%%%

%%%%%%%%%%%%%%%%%%%%%%%%%%%%%%%%%%%%%%%%%%%%%%%%%%%%%%%%%%%%%%%%%

\section{Results}\label{sec:Res}

\begin{figure*}
	\includegraphics[trim={0 0 0 0},clip]{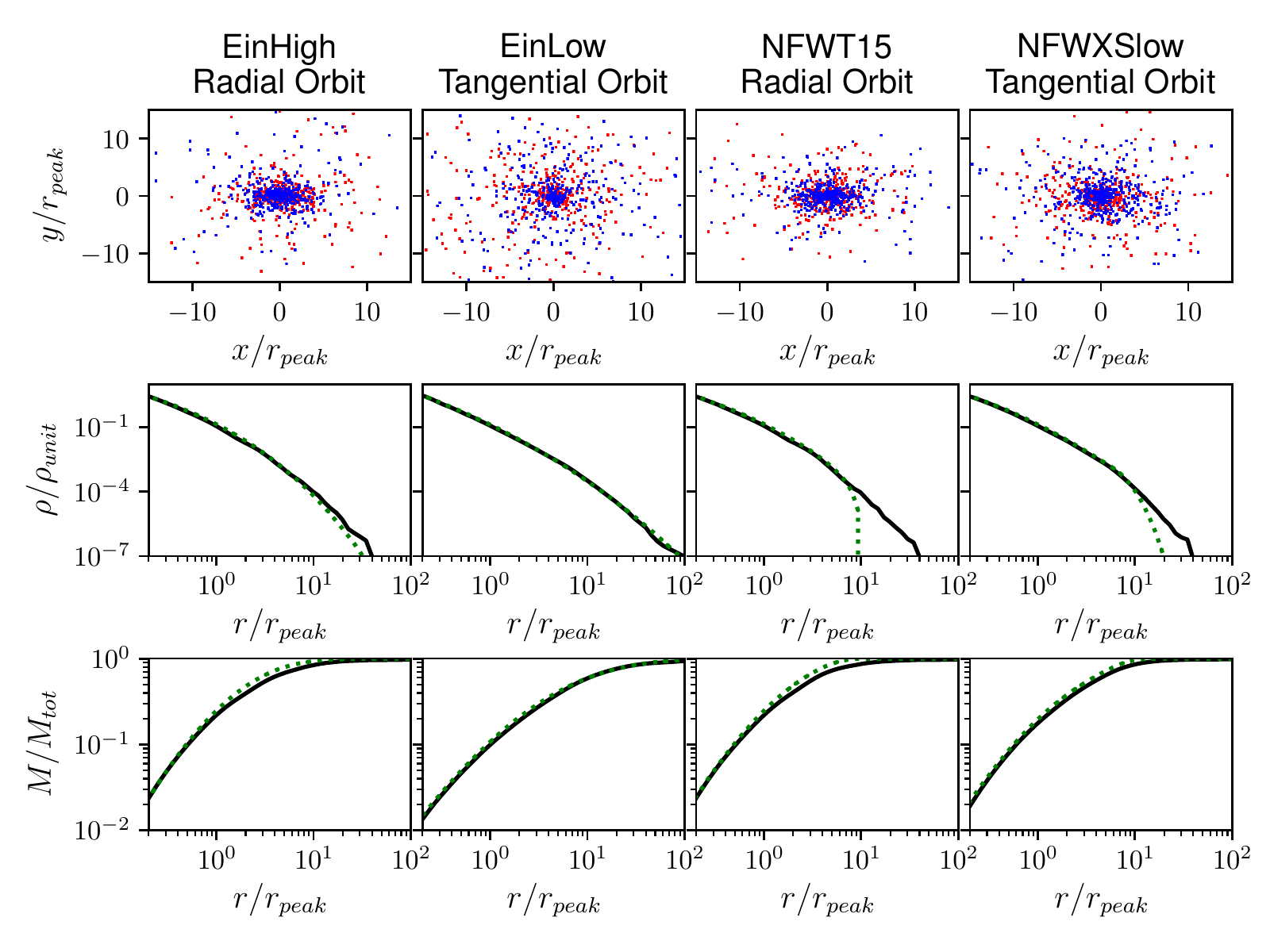}.
	\caption{Sample results from four merger simulations. The top row of panels shows the state of the remnant at $t= 300\, t_{\rm unit}$ (in this plot we show only $10^3$ randomly selected particles). The particles are coloured either red or blue, depending on their initial halo. The middle row of panels shows the density profiles, and the bottom row shows the mass profiles. The scaled ICs and remnants are shown with dotted green lines and solid black lines, respectively. All haloes were initially separated by $r_{\rm sep} = 10$, and given either a radial or tangential velocity of $v_0 = 0.8 v_{\rm esc}$, where $v_{\rm esc}$ is the escape speed of a point mass in the potential of the initial halo.}
	\label{fig:Overview}
\end{figure*}

 Fig.~\ref{fig:Overview} shows sample results from four different merger simulations. The top panels show the resulting remnants (using a random subset of $10^3$ particles), the middle panels show the density profiles, and the bottom panels show the cumulative mass profiles. In general, we find that the remnants are non-spherical in shape, with the radial and tangential orbits producing prolate and oblate systems, respectively. We can compare the final density and mass profiles (solid black lines) to those of the initial haloes (dotted green lines), scaling the mass by a factor of $2$ and the radius by a factor of $2^{1/3}$, as expected if the remnant has the same mean density as the ICs. We see some differences between the remnant and scaled ICs; specifically, the density profile changes slightly in curvature, and there appears to be some mass rearrangement. In general the remnant profile appears to be more extended than the ICs, but there is also a slight increase in central density. The density profiles of the remnants will be considered in detail in the next paper in this series.

%%%%%%FIGURE 10%%%%%%

\subsection{Scaled energy and angular momentum}

Changes in halo structure should presumably depend not on the total orbital energy, but on the fraction of this energy that is available as internal energy, given momentum conservation requires some bulk motion of the remnant. Before the merger, the total energy of the system is equal to the internal energy of the two initial haloes, plus the initial orbital energy: 
\begin{equation} 
E_{\rm tot} = 2E_0 + E_{\rm orb}\, . 
\end{equation}
After the merger, the remnant will have  internal energy $E_0'$, plus some net kinetic energy, $K_{\rm orb}'$; $E_{\rm tot}' = E_0' + K_{\rm orb}'$. The net orbital kinetic energy, $K_{\rm orb}'$, can be calculated from conservation of momentum, $K_{\rm orb}' = K_{\rm orb} M/M'$, where $K_{\rm orb}$ is the initial (orbital) kinetic energy, $M$ is the mass of the initial halo, and $M'$ is the mass of the remnant. For an equal mass merger, $K_{\rm orb} = 1/2 M v_0^2$, and $M/M' = 1/2$. Thus, since $E_{\rm tot} = E'_{\rm tot}$, the internal energy of the remnant halo will be:
\begin{equation} \label{eq:E0p}
E_0' = E_{\rm orb}+2 E_0 - \dfrac{1}{4} M v_0^2 \,\,\,.
\end{equation}
We found that calculating $E_0'$ in this way agrees to within 2 per cent with a direct calculation of the internal energy of the remnant using equation~\eqref{eq:KandP}.

In a cosmological context, encounters between haloes may be close to parabolic, with $E_{\rm orb} \sim 0$ in the centre-of-mass frame. In this case, the total energy of the remnant is then just twice the energy of the initial haloes. If the form of the density profile is conserved, from the scaling of potential energy we expect that the size of the remnant will increase linearly with mass, i.e. by a factor of 2 \citep{farouki1983}. As a result, the density of the remnant will be lower than that of the initial haloes. This evolution at constant specific energy has been considered the baseline in previous studies of mergers \citep[e.g.][]{navarro1989}. On the other hand, we are interested in following the evolution of shape and concentration in mergers partly to determine how the mean density of structures evolves with time, for instance in order to calculate the boost factor for dark matter annihilation \citep[e.g.][]{okoli2018}, which scales as $\int \rho \, \rm d m$. Thus, we will consider the baseline case one in which the {\it overall density distribution} is conserved. We define `self-similar evolution' to be evolution where the relative mass fraction at any given density in a structure remains constant, and thus the mean density and boost factor do not change.

In the case of self-similar evolution, $r \sim M^{1/3}$ and thus $E_0 \propto M^{5/3}$. Thus in the self-similar case for equal-mass mergers, the internal energy of the halo should increase by $2^{5/3}$. More generally,  we can define a new parameter, the change in internal energy relative to the change expected in self-similar evolution:
\begin{equation}\label{eq:kapp_def}
\kappa \equiv \dfrac{E_0'}{E_0} \left(\dfrac{M}{M'}  \right)^{5/3} \,,
\end{equation}
This parameter provides a convenient dimensionless  measure of the change in internal energy of the halo; a value of $\kappa=1$ corresponds to a self-similar change in energy. If $\kappa <1$, then the remnant is less bound than the progenitor, while if $\kappa >1$ the remnant is more bound.

We can also express angular momentum using a dimensionless spin parameter. The spin parameter was originally defined by \cite{peebles1971} as
\begin{equation} \label{eq:lam}
\lambda= \dfrac{\sqrt{|E_0|}|\textbf{J}|}{GM^{5/2}} \,\,\, ,
\end{equation}
while an alternative definition commonly used in the literature was proposed by \cite{bullock2001}:
\begin{equation}
\lambda_{\rm B} = \dfrac{|\textbf{J}|}{\sqrt{2}M_{\rm vir} r_{\rm vir}v_{\rm vir}} \,\,\,.
\end{equation}
The second definition, $\lambda_{\rm B}$, is equivalent to $\lambda$ under the assumption of virialization and an isothermal profile, and is often preferred in the literature since calculating mass is much easier than calculating the full energy. However, these assumptions lead to scatter in the expected spin \citep{ahn2014}. Additionally, $\lambda_B$ is not well-defined in non-cosmological simulations, and therefore we will use $\lambda$ as defined in equation~\eqref{eq:lam}.

\begin{figure}
	\includegraphics[width = \columnwidth,trim={3cm 5cm 4cm 0},clip=true]{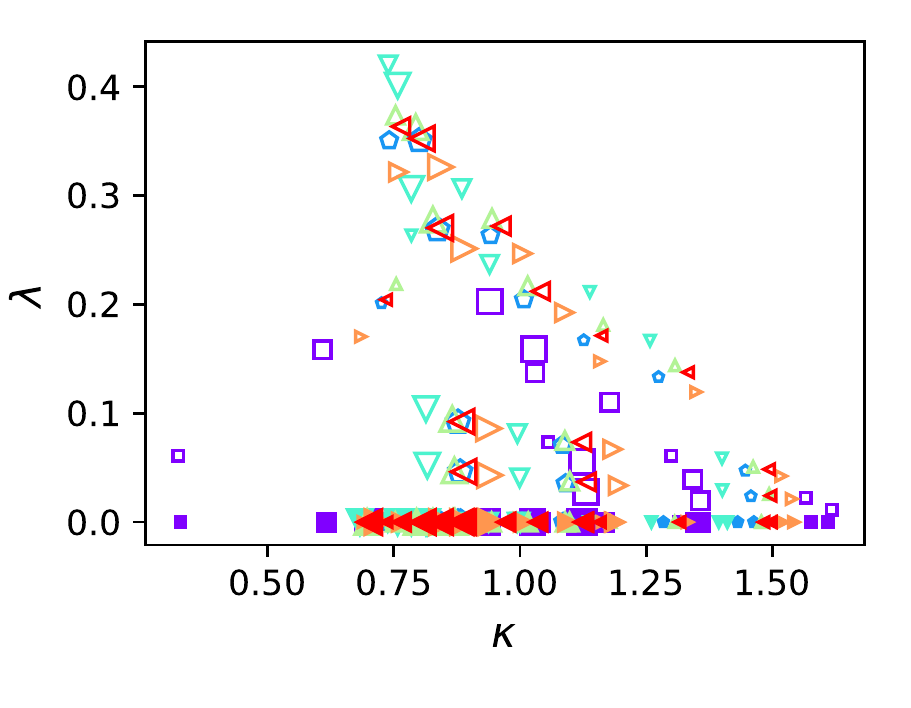}.
	\caption{Dimensionless energy and spin parameters $\kappa$ and $\lambda$, for the full set of simulations. Symbols are as in Fig.~\ref{fig:OrbitalParameters}.
	}	
	\label{fig:OrbitalParameters2}
\end{figure}

The expected spin parameter of the remnant can be predicted from the ICs, using equations~\eqref{eq:L} and \eqref{eq:E0p}. Comparing this to a direct calculation of the spin parameter of the remnant, we find the two agree to within $1$ per cent. The orbital parameters $\kappa$ and $\lambda$ for all our simulations are shown in Fig.~\ref{fig:OrbitalParameters2}. We note that there are restrictions in this parameter space in low-$\kappa$/high-$\lambda$ as well as in high-$\kappa$/high-$\lambda$ regions. The former is because orbits become unbound as $\kappa$ increases, while the latter is due to the amount of free energy available.

%%%%%%FIGURE 11%%%%%%

\begin{figure}
	\includegraphics[width = \columnwidth,trim={0cm 5cm 6cm 0},clip]{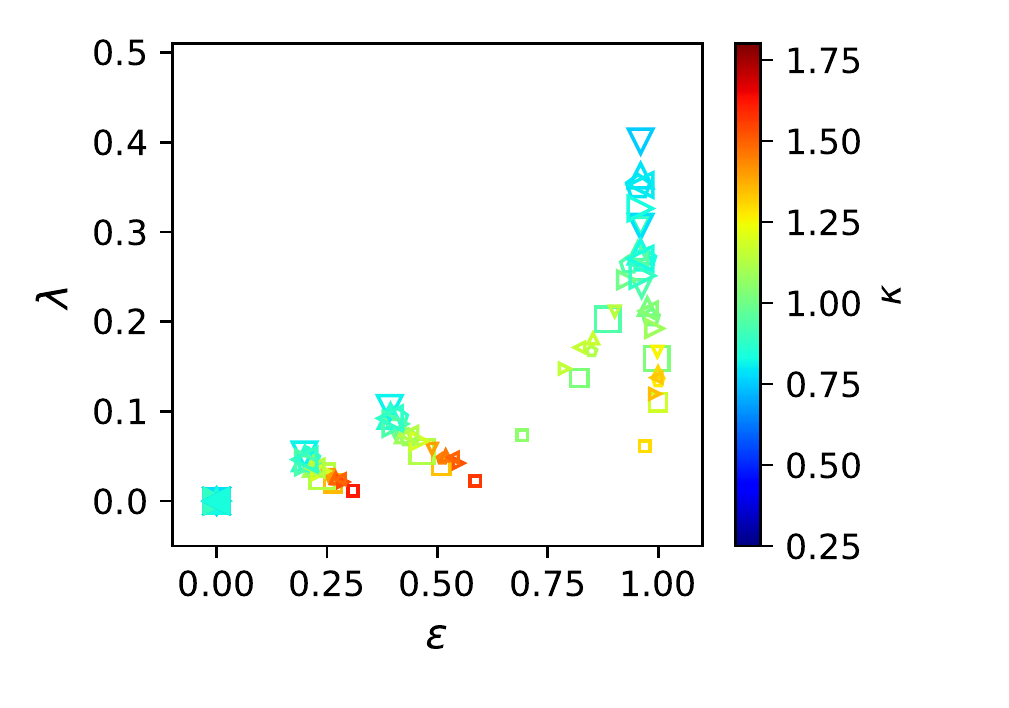}.
	\caption{ Spin parameter of the merger remnant, $\lambda$, versus the circularity of the initial orbit, $\epsilon$. 
	Symbols are as in Fig.~\ref{fig:OrbitalParameters}.}
	\label{fig:Spin_vs_Orbit}
\end{figure}

Since $\epsilon$ and $\lambda$ are both measures of the angular momentum of the initial orbit, they will clearly be correlated. Fig.~\ref{fig:Spin_vs_Orbit} compares the value of the two parameters, to clarify this relationship. Circularities are calculated assuming 
the orbital energy of the second halo is that of a point mass orbiting in the potential of the first halo. Note that circularity cannot be 
calculated for the higher energy orbits (since there is no bound circular orbit with the same energy), and therefore this plot contains 
only a subset of the simulations. Although $\epsilon$ and $\lambda$ are (positively) correlated, there is also an energy dependence 
in both definitions; for the same angular momentum, spin increases with kinetic energy (and thus decreases with increasing $\kappa$).

%%%%%%FIGURE 12%%%%%%

\subsection{Halo alignment}
	
\begin{figure*}
	\centering
	\subfloat{{\includegraphics[trim={0cm 1.5cm 0cm 1.5cm},clip]{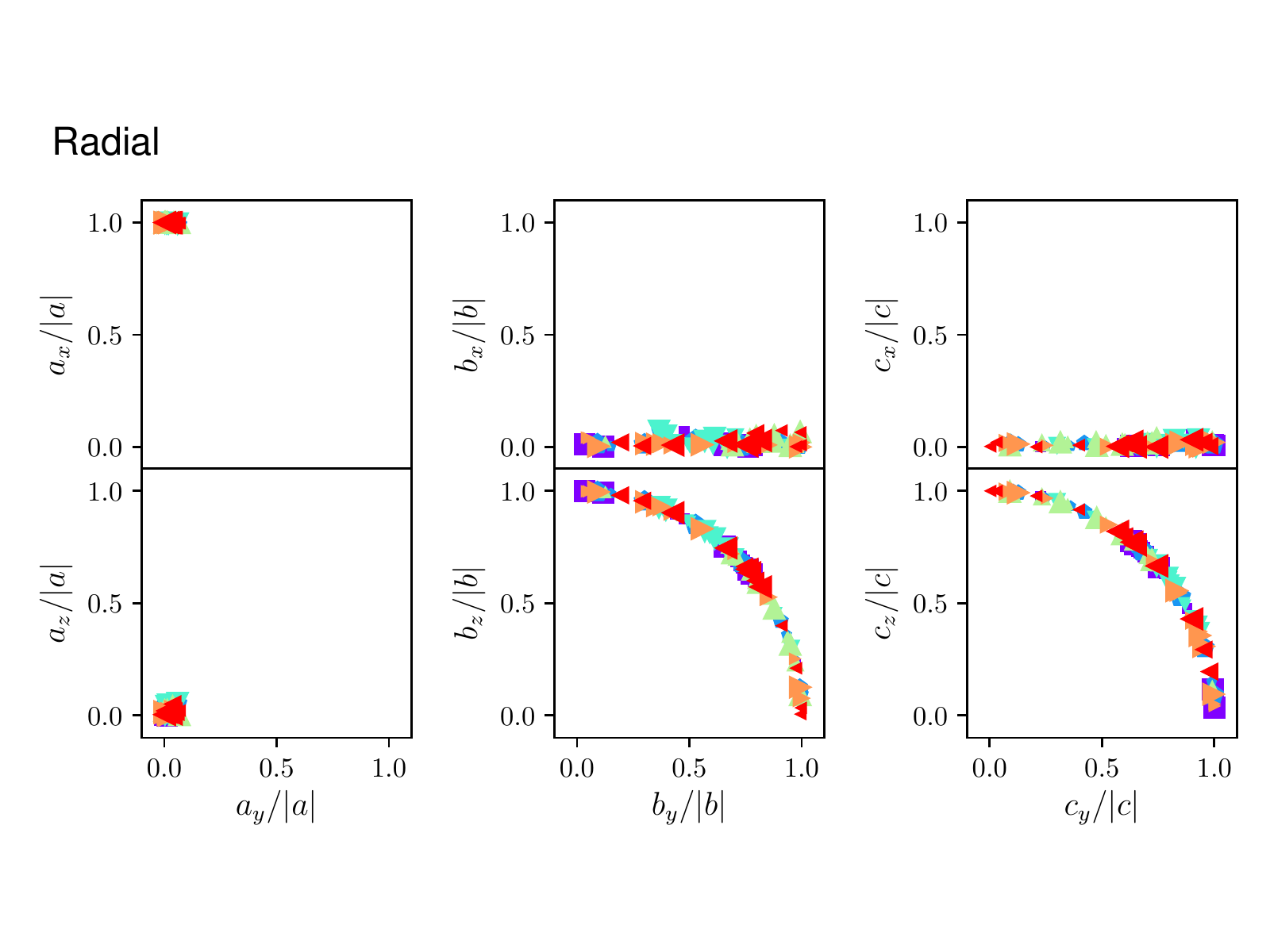}}}%

	\subfloat{{\includegraphics[trim={0cm 1.5cm 0cm 1.5cm},clip]{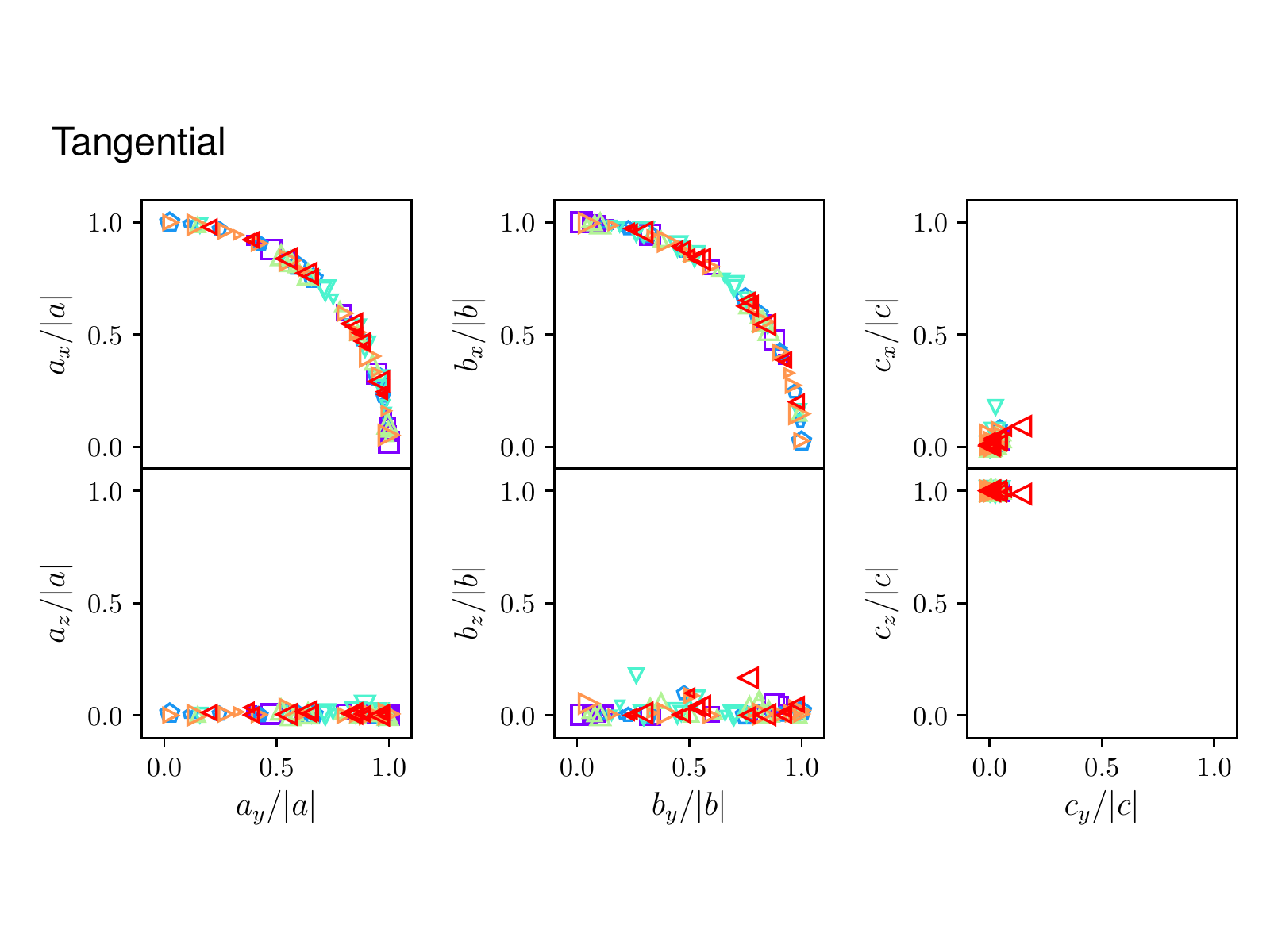}}}%
	\caption{Normalized $x$, $y$, and $z$-components of the normalized principal axes $a$, $b$, and $c$ of the remnants of halo mergers with either (top) radial or (bottom) tangential initial velocity, at the final time. Colours and symbols are as in Fig.~\ref{fig:OrbitalParameters}.}
	\label{fig:RotationAxes}
\end{figure*}

Fig.~\ref{fig:RotationAxes} shows the final alignment of the halo remnants. Haloes were initially separated along the $x$-axis, and given an initial velocity in the $x$-axis (radial orbits) or in the $y$-axis (tangential orbits). Fig.~\ref{fig:RotationAxes} demonstrates that haloes on radial orbits have their major axis, $a$, aligned along the axis of the merger, $x$. The two other axes, $b$ and $c$, lie in the $y$-$z$ plane. On the other hand, for tangential orbits, axes $a$ and $b$ lie in the $x$-$y$ plane, while the minor axis $c$ points in the $z$ direction. Clearly, the shape of the remnant is aligned with the direction of the merger as expected from previous work \citep[e.g.][]{maccio2007,veraciro2011}. Since it has been suggested that radial orbits produce prolate haloes, while tangential orbitals produce oblate haloes \citep[e.g.][]{moore2004}, there seems to be a link between halo orientation and shape. The effects of orbital parameters on halo shape will be explored further in Section~\ref{sec:shape}.

\subsection{Net change in halo size}

\begin{figure}
	\includegraphics[width = \columnwidth,trim={0cm 5.5cm 6cm 0cm},clip]{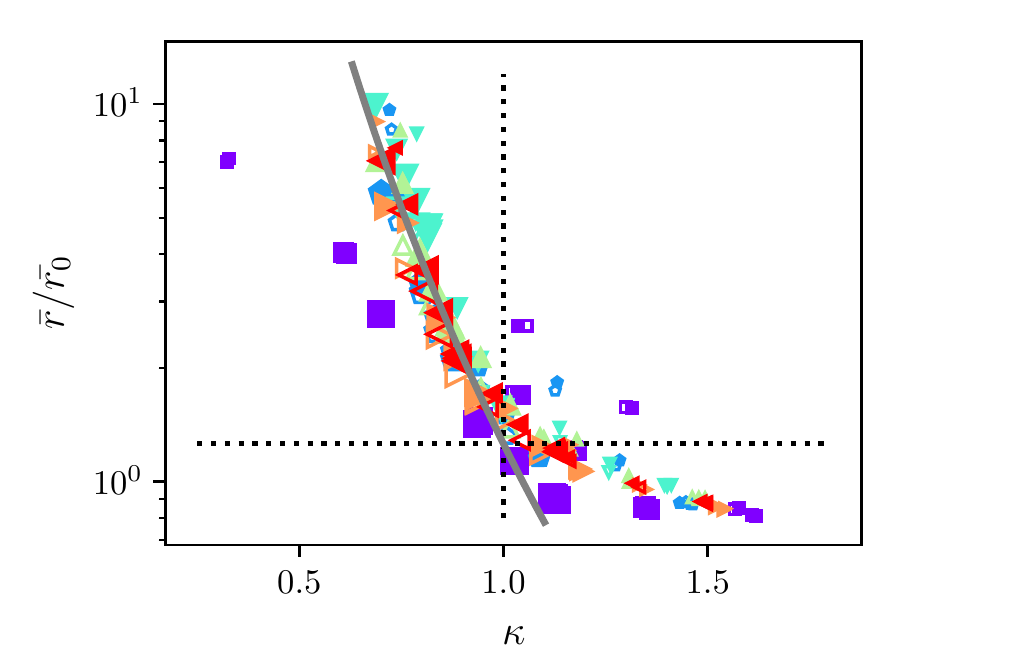}.
	\caption{The mean radial extent of the remnant, $\bar{r}$, relative to the mean radial extent the initial halo models, $\bar{r}_0$, versus the energy parameter $\kappa$. The self-similar expectations are shown with dotted black lines. The solid gray line is when $\bar{r}/\bar{r_0} = 2^{1/3} \kappa^{-5}$.
	Symbols and colours are as in Fig.~\ref{fig:OrbitalParameters}.}
	\label{fig:R_mean}
\end{figure}

After the merger, we expect the remnant to be larger than either of the initial haloes, and possibly also elongated in the merger direction, at least in the case of more radial mergers. As a profile-independent measure of size, we define the `radial extent' of a system (either the merger remnant, or an initial halo) to be the mean distance of all particles in the halo from the centre of the system, where the latter was determined as discussed in Section~\ref{sec:SetUp}. Extents along the principal axes are defined similarly, as the mean distance projected on each axis. In Fig.~\ref{fig:R_mean} we show the radial extent of the merger relative to the radial extent of the ICs, as a function of $\kappa$. The size of the remnant, relative to the ICs, depends mainly on $\kappa$, though there is also a small dependence on the  initial halo model. The EinLow simulations (squares) do not increase in size as much as the other initial halo models for high-energy (low $\kappa$) orbits. This may be because the EinLow ICs are very extended compared to the other models. The results go through the self-similar expectation for an equal mass merger, $\bar{r}/\bar{r}_0 = 2^{1/3}$ when $\kappa=1$. If $\kappa>1$ (more bound remnants), the remnant is smaller than expected in the self-similar case, and may even have a radial extent smaller than that of the initial haloes. If $\kappa>1$ (less bound remnants), the remnant is larger than expected from self-similar scaling.  There is little or no dependence on whether the orbit is radial or tangential, nor on the parameter $r_{\rm sep}$.

For $\kappa<1$, we find that $\bar{r}/\bar{r}_0 \approx 2^{1/3} \kappa^{-5}$; this might be expected since the energy of a virialized self-similar halo scales as $M^2/r\sim r^5$, and therefore the change in halo radius scales as $\kappa^{-5}$. However, as the size of the halo decreases, the dependence on  $\kappa$ weakens; this may be because for these very low-energy orbits, the halo remnant is not self-similar to the ICs. The EinLow simulations  also have a weaker dependence on $\kappa$, and this may once again indicate departures from self-similarity. We will explore the self-similarity between initial halo models and final  remnants in a companion paper, \citep{drakos2018}.

%%%%%%FIGURE 13%%%%%%

\begin{figure*}
	\includegraphics[trim={0cm 0 0cm 0},clip]{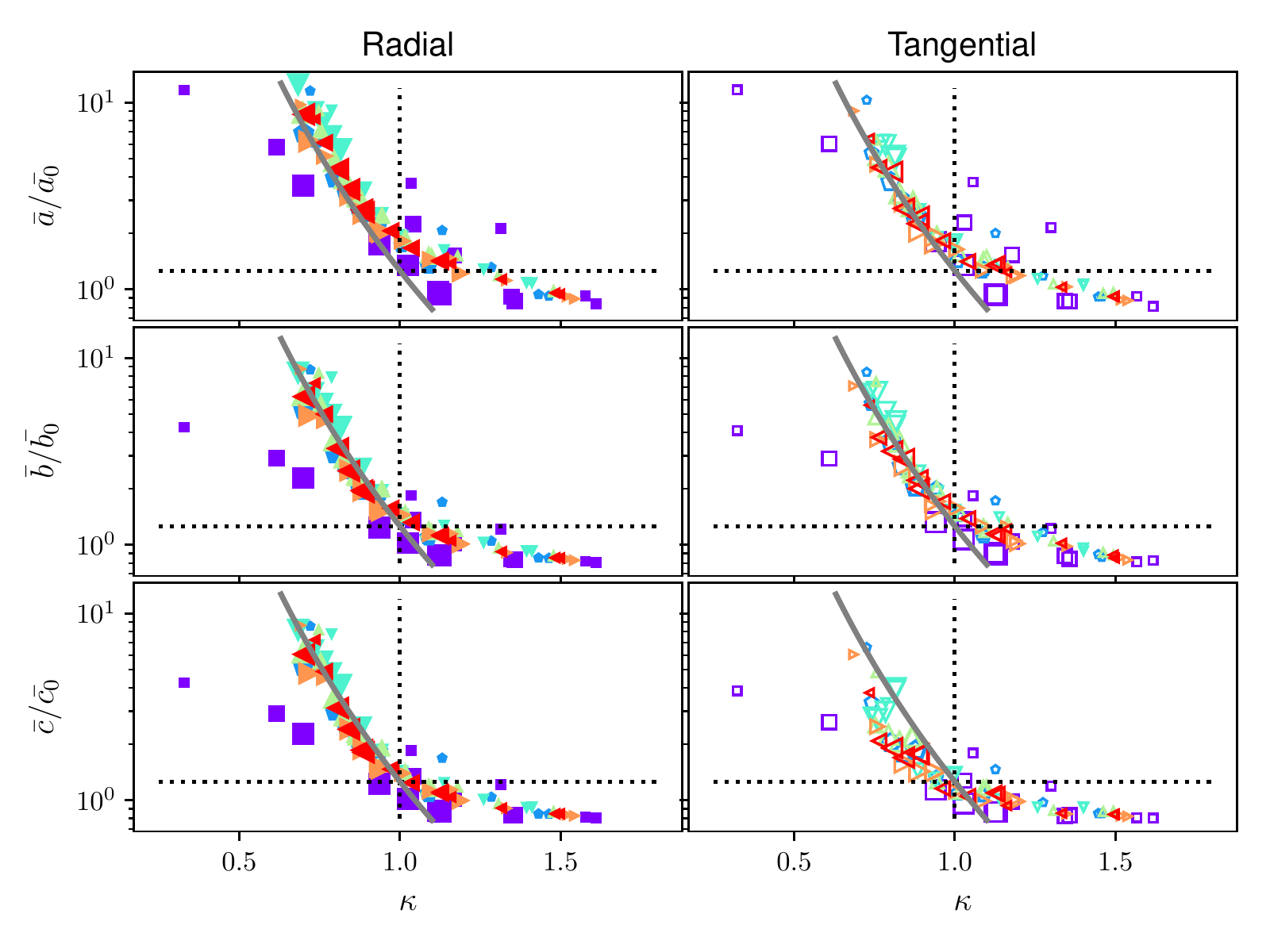}.
	\caption{The change in the mean extent along the three principal axes, $a$, $b$, and $c$, versus $\kappa$. The solid gray line is when the change in the mean extent is equal to $2^{1/3} \kappa^{-5}$.
	The self-similar expectations are shown with the dotted black lines. Symbols and colours are as in Fig.~\ref{fig:OrbitalParameters}.  The left- and right-hand panels are simulations with radial and tangential initial velocities, respectively.}
	\label{fig:Ax_mean}
\end{figure*}

Similar trends can be found when comparing the mean extent of the remnant projected along each of the principal axes (Fig.~\ref{fig:Ax_mean}). Relative to the initial halo models, the size of the remnant is largest along the major principal axis $a$ by definition. The extent of the halo along this axis increases slightly more than expected from self-similar scaling when $\kappa=1$. We might expect the intermediate axis of the remnant, $b$, to be larger for tangential orbits than for radial orbits, because the orbit lies in the $a$-$b$ plane; it seems, however, that $b$ changes by roughly the same amount in radial and tangential cases, but that the minor axis $c$ (perpendicular to the orbital plane) is smaller in tangential cases compared to radial cases.

%%%%%%FIGURE 14%%%%%%

We can derive a simple theoretical prediction for the expected size of the remnants along the major axis. In the spherical collapse model, when a cosmological overdensity collapses and virializes, the final radius of each shell is equal to half its radius at turnaround, as a consequence of energy conservation and the (scalar) virial theorem. By analogy, if the merger remnants in our simulations were to conserve the virial tensor component-wise, we might expect their extent along the major axis to be half the turnaround radius of the initial two-halo system, $r_{\rm TA}$. Since the virial radius should also increase by a factor of $2^{1/3}$ due to the extra mass in the system, however, we expect the virial radius along the longest axis to be $r_{\rm vir} \approx (r_{\rm TA}/2)2^{1/3} = r_{\rm TA}/2^{2/3}$. 

\begin{figure}
	\includegraphics[width = \columnwidth,trim={0.5cm 5.5cm 8cm 0},clip]{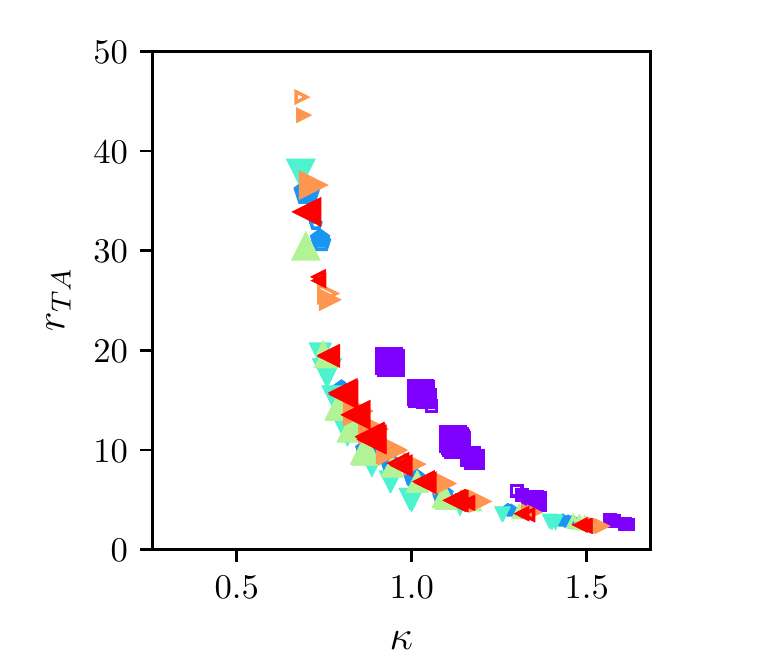}.
	\caption{Turnaround radius as a function of $\kappa$. Symbols and colours are as in Fig.~\ref{fig:OrbitalParameters}.}
	\label{fig:TurnAround_vs_Energy}
\end{figure}

To determine the turnaround radius, we calculated $P_{\rm orb}(r)$ for each set of ICs by placing the two initial haloes a distance of $r$ apart and calculating $P_{\rm orb} = P - 2P_0$, where the total potential $P$ and internal potential energy of each halo, $P_0$, were calculated from equations~\eqref{eq:KandP} and \eqref{eq:P_0}, respectively. The turnaround radius is then the radius such that $E_{\rm orb} = P_{\rm orb}(r_{\rm TA})$; i.e.~the radius at which there is no kinetic orbital energy. This was determined by first smoothing $P_{\rm orb}(r)$, using a Gaussian filter, and then interpolating this smoothed potential to find $r_{\rm TA}$.  
Fig.~\ref{fig:TurnAround_vs_Energy} shows the turnaround radius as a function of $\kappa$. For high-energy orbits with small $\kappa$
the turnaround radius is very large, but then it goes to zero for large $\kappa$.

%%%%%%FIGURE 15%%%%%%

\begin{figure}
	\includegraphics[width = \columnwidth,trim={3cm 0cm 4cm 0},clip]{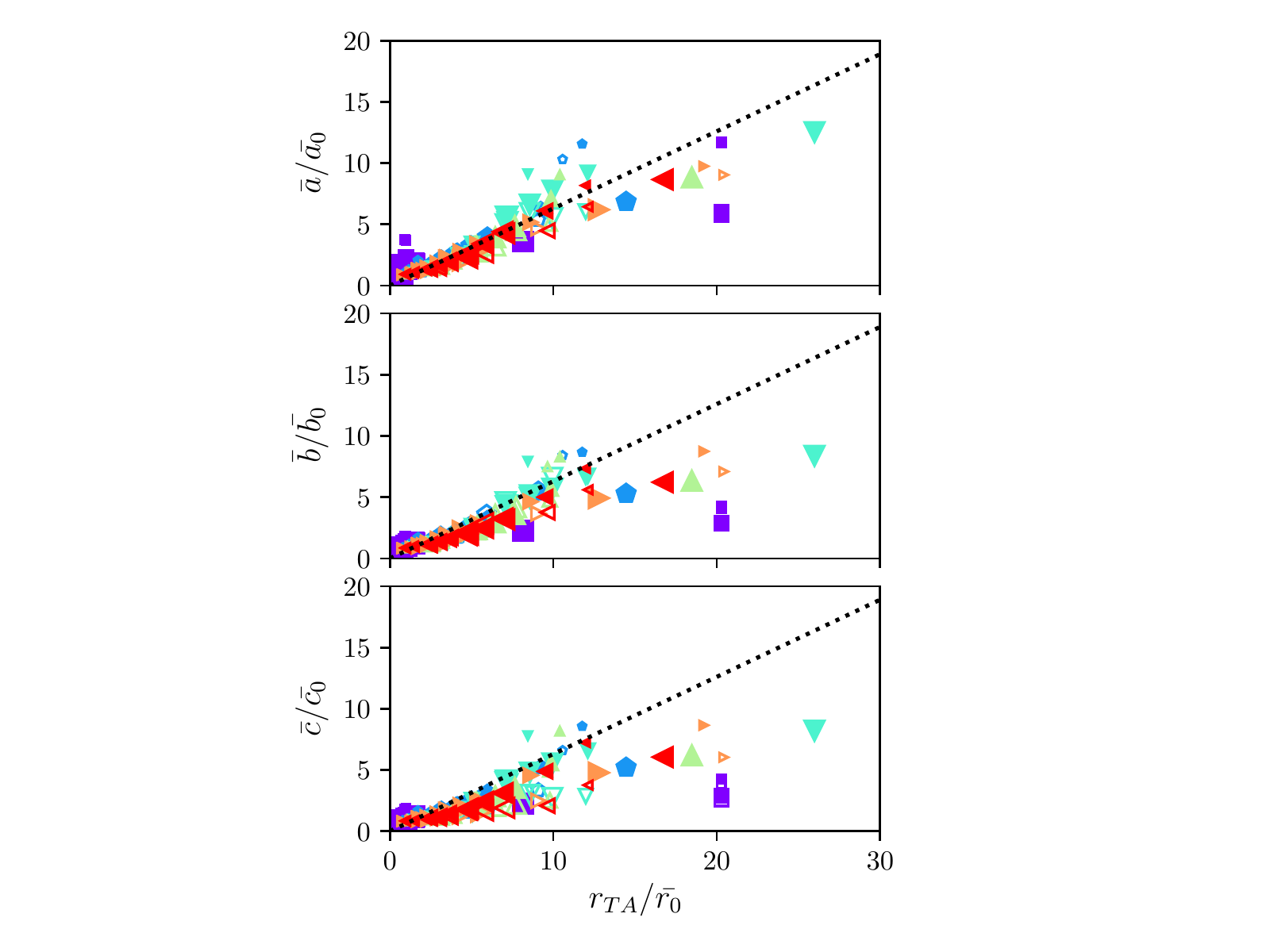}.
	\caption{The size of the remnant versus the turnaround radius of the initial orbit. The size of the remnant is measured by the mean extent along the three principal axes, $a$, $b$ and $c$, and the turnaround radius is normalized by the average radial extent of the initial halo. 
	Dotted lines are the prediction from equation~\eqref{eq:a_predict}. Symbols and colours are as in Fig.~\ref{fig:OrbitalParameters}.}
	\label{fig:TurnAround_size}
\end{figure}

To test the prediction that $r_{\rm vir}' \approx r_{\rm TA}/2^{2/3}$ along the largest axis, we assume that the virial radius scales as the average particle distance in the ICs, $\bar{r}_{0}$, and propose:
\begin{equation} \label{eq:a_predict}
\dfrac{\bar{a}}{\bar{a}_0} =  2^{-2/3} \dfrac{r_{\rm TA}}{\bar{r}_0} \,\,\, .
\end{equation}
We compare the change in all three principal axes in Fig.~\ref{fig:TurnAround_size}. We find the extent along the largest axis, $\bar{a}$, does indeed scale as predicted, albeit with considerable scatter for large values of $r_{\rm TA}$. Since $r_{\rm TA}$ decreases exponentially with $\kappa$, the larger $r_{\rm TA}$ values are more sensitive to the interpolation used to estimate them, so this effect may be purely numerical. For the low energy orbits (high $\kappa$, low $r_{\rm TA}$), the extents along the other axes $b$ and $c$ behave similarly to $a$, but for higher energy orbits, the change is smaller than our prediction.
 
%%%%%%FIGURE 16%%%%%%

\subsection{Net change in halo shape} \label{sec:shape}

\begin{figure*}
	\includegraphics[trim={0cm 0cm 0cm 0},clip]{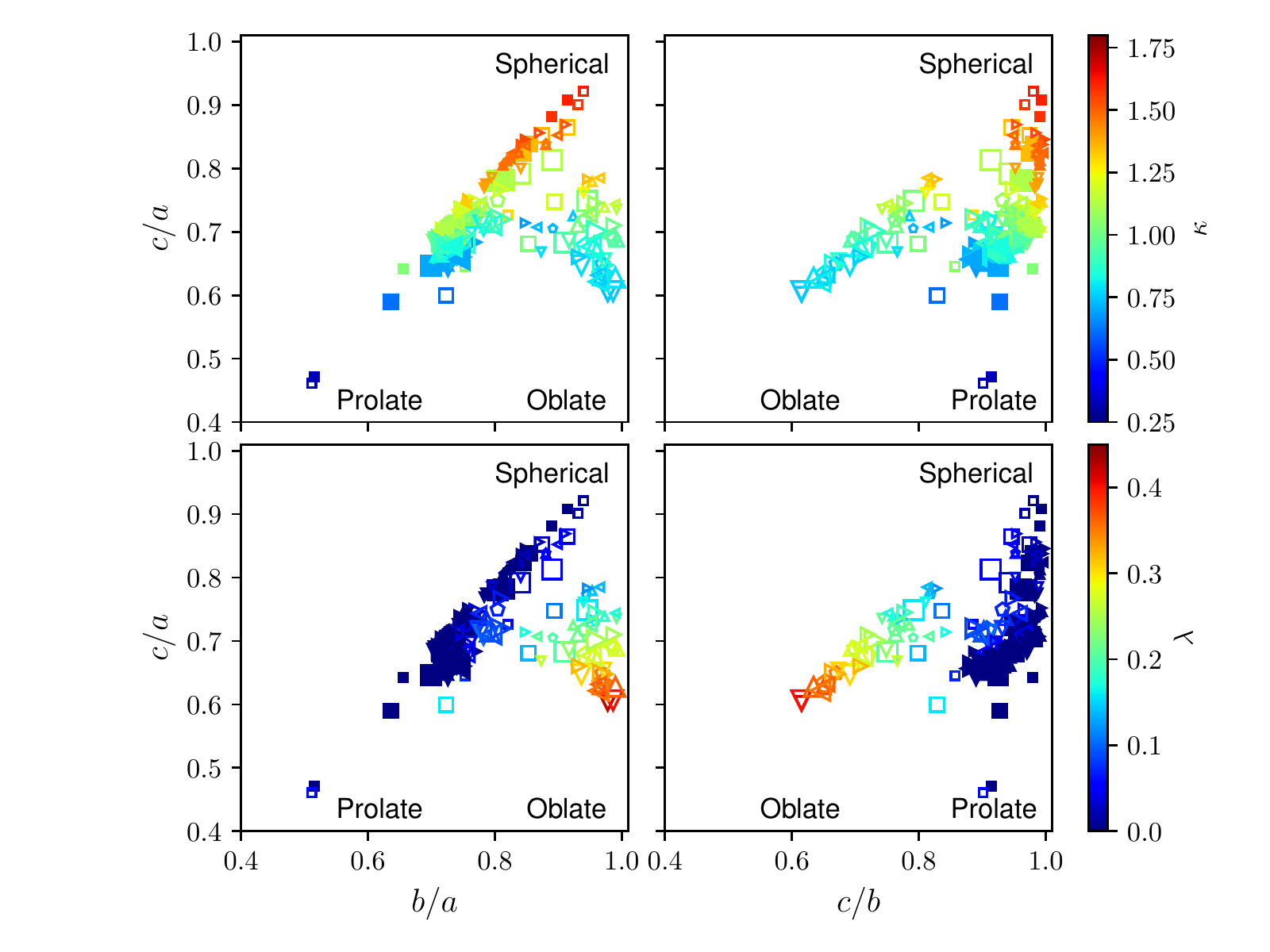}.
	\caption{Remnant axis ratios $c/a$ vs $b/a$ (left) and $c/a$ vs $c/b$ (right), where $a>b>c$. Regions of parameter space corresponding to spherical, prolate and oblate haloes are labeled. The top panels are coloured by the relative energy parameter $\kappa$, and the bottom panels by the spin parameter, $\lambda$. Symbols are as in Fig.~\ref{fig:OrbitalParameters}.}
	\label{fig:Shape}
\end{figure*}

The axis ratios of the merger remnants were calculated as in Section~\ref{sec:shape_measure} and are shown in Fig.~\ref{fig:Shape}. We emphasize that these measurements of principal axes sizes, $a$, $b$, and $c$ are not the same as the extent,  $\bar{a}$, $\bar{b}$, and $\bar{c}$, discussed in the previous section, which were calculated as the mean particle distance projected along the principal axes. The top and bottom panels of Fig.~\ref{fig:Shape} are coloured by the relative energy parameter $\kappa$ and spin parameter $\lambda$, respectively. Generally speaking, more bound remnants are also more spherical. The shape ratio $c/a$ depends mainly on energy, and is smaller when $\kappa$ is smaller (less bound haloes). The parameter $c/b$ depends mainly on $\lambda$, and $b/a$ depends on both $\kappa$ and $\lambda$. There is little or no dependence on the IC models or on the parameter $r_{\rm sep}$ on the final shape of the remnant. Spin dictates whether the final remnant is prolate or oblate in shape; mergers on (low-spin) radial orbits produce prolate haloes with $c/b \approx 1$, while mergers on tangential (high-spin) orbits produce oblate haloes with $c/b \approx c/a$. For the largest spin values considered here, $c/b \approx c/a \approx 0.6$. Overall, these results are consistent with the two cases considered in \cite{moore2004}.

%%%%%%FIGURE 18%%%%%%

\begin{figure}
	\includegraphics[width = \columnwidth,trim={6cm 6.5  0.5cm 0},clip]{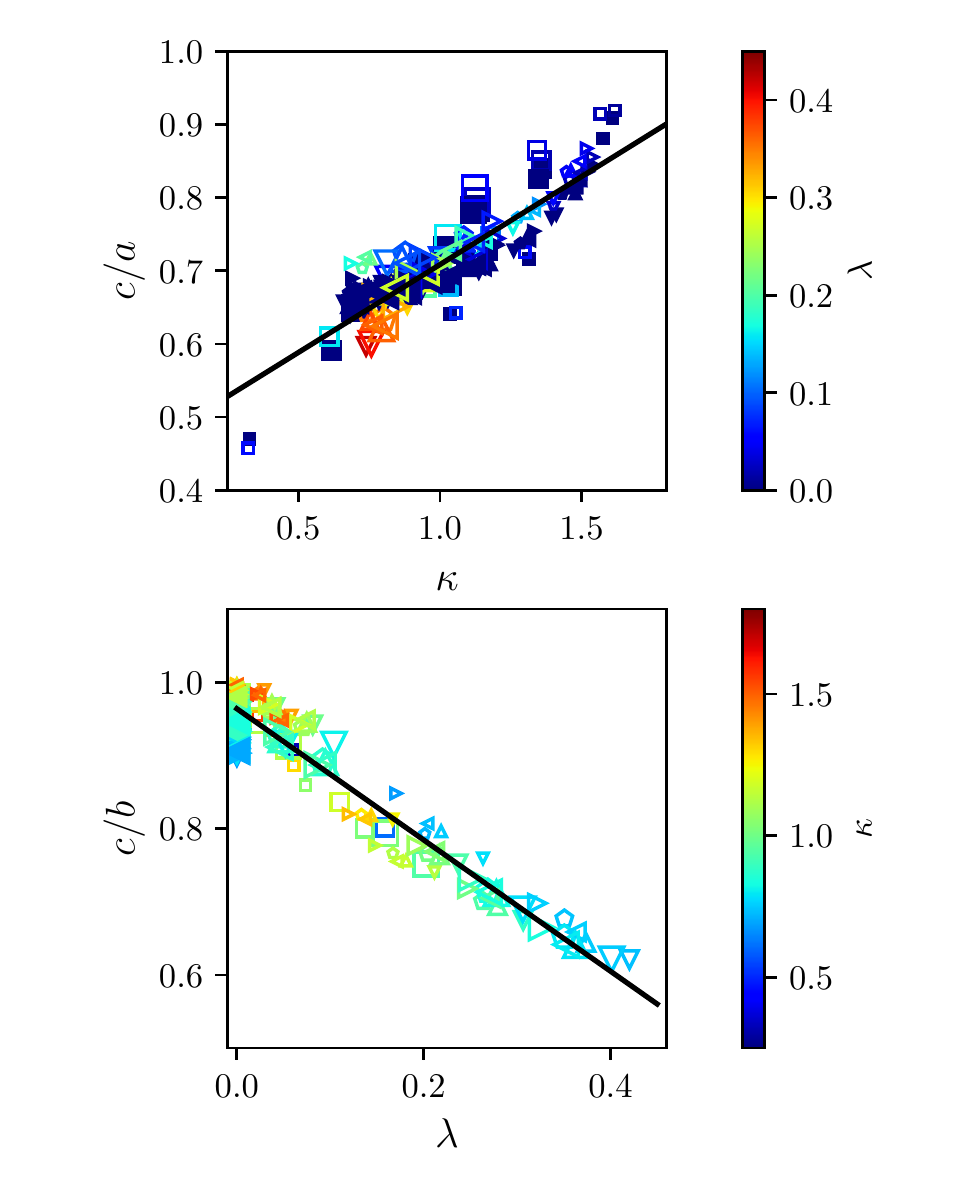}.
	\caption{Remnant axis ratios $c/a$ (top) and $c/b$ (bottom) as a function of the relative energy parameter $\kappa$, and the spin parameter, $\lambda$, respectively. Symbols are as in Fig.~\ref{fig:OrbitalParameters}. Fits to the average trends are given in the text; the RMS scatter with respect to each is 
		approximately 0.03.}
	\label{fig:Shape_vs_Orbit}
\end{figure}

Fig.~\ref{fig:Shape_vs_Orbit} shows the final shape parameters, $c/a$ (top) and $c/b$ (bottom) of the halo remnants as a function of $\kappa$ and $\lambda$, respectively. We also show fits to the main trends:
\begin{subequations} 
\begin{align}
\label{eq:shapesca}
c/a &= 0.24 \kappa+ 0.47\\ 
\label{eq:shapescb}
c/b&= -0.90 \lambda + 0.96 \,\,\,. 
\end{align}
\end{subequations}

\begin{figure}
	\includegraphics[width = \columnwidth,trim={6cm 6.5  0.5cm 0},clip]{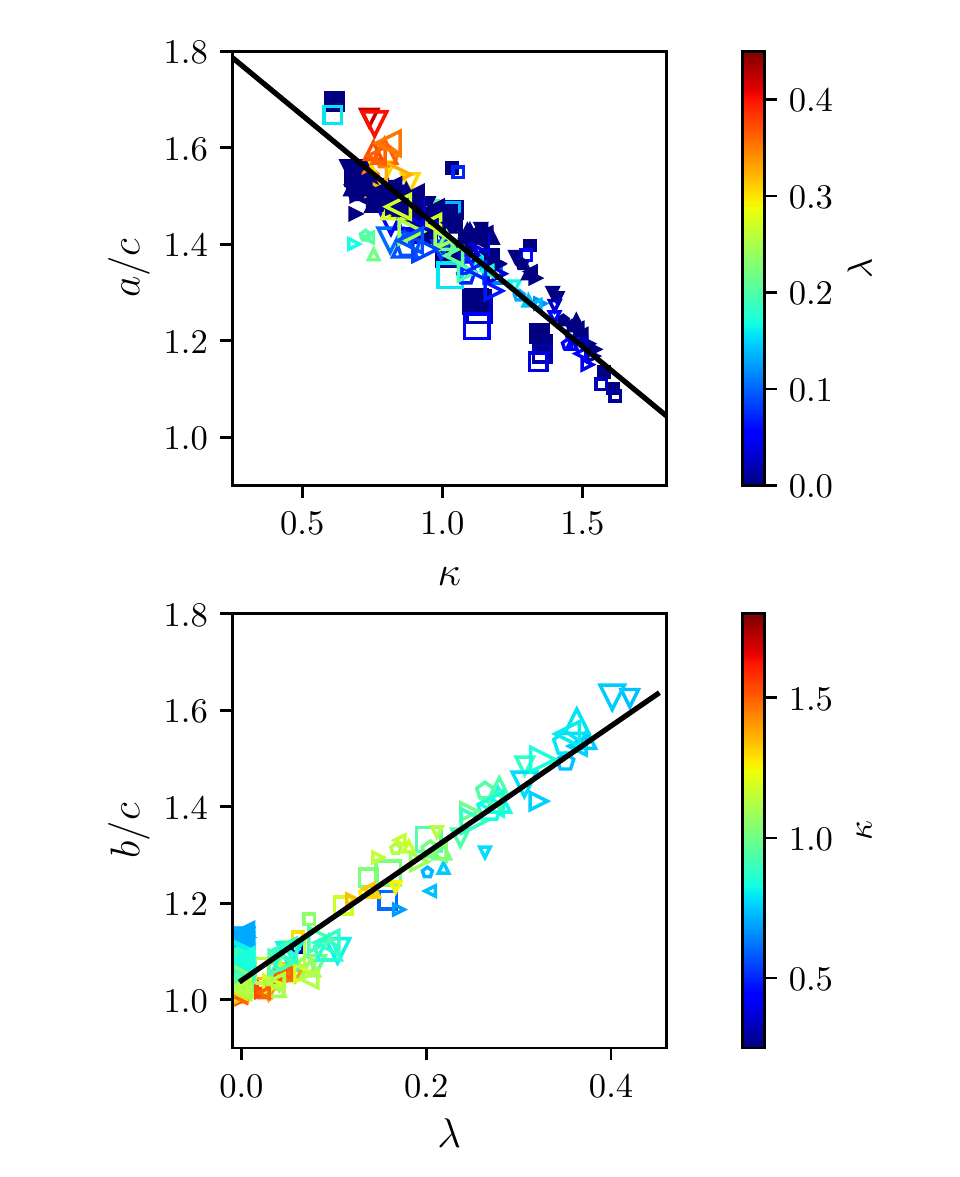}.
	\caption{Remnant axis ratios $a/c$ (top) and $b/c$ (bottom) as a function of the relative energy parameter $\kappa$, and the spin parameter, $\lambda$, respectively. Symbols are as in Fig.~\ref{fig:OrbitalParameters}. Fits to the average trends are given in the text; the RMS scatter with respect to each is approximately 0.07 (top) and 0.04 (bottom).}
	\label{fig:Shape_vs_Orbitv2}
\end{figure}

Alternatively, we could fit the inverse ratios $a/c$ and $b/c$, since the axis $c$ generally grows less than $a$ or $b$ after the merger. Fig.~\ref{fig:Shape_vs_Orbitv2} shows these ratios, along with the fits:
\begin{subequations} 
	\begin{align}
	\label{eq:shapesac}
	a/c &= -0.50 \kappa+ 1.91\\ 
	\label{eq:shapesba}
	b/c&= 1.32 \lambda + 1.03 \,\,\,. 
	\end{align}
\end{subequations}

In the bottom panel of Fig.~\ref{fig:Shape_vs_Orbit}, it is a bit surprising that $c/b \neq 1$ in some cases, even when $\lambda=0$. As discussed previously, by symmetry, one would expect $c = b$ for mergers on purely radial orbits. It seems likely that $c \neq b$ in practice for numerical reasons arising from the fact that the ICs are not perfectly symmetric. We expect this to be particularly true at high energies, where the simulation is sensitive to the direction of the initial velocity. Any noise or uncertainty in the shape measurement will result in an underestimate of $c/b$, since by definition $c < b$, and thus $c/b<1$. 

%%%%%%FIGURE 19%%%%%%

One implication of equation~\eqref{eq:shapesca} is that for a self-similar change in energy, $c/a \approx 0.7$. To understand this, we can consider a self-similar radial merger between two spherical, equal-mass haloes. In the self-similar case, the radius of the remnant should scale as $r'/r = 2^{1/3}$. Further, we will assume that the original size of the halo is $x_0$ along any of the principal axes, and that only the $a$ axis increases in size, such that $a'= \beta x_0$, $b'=x_0$, and $c'=x_0$. Then,
\begin{equation} \label{eq:beta}
\begin{aligned}
\dfrac{r'}{r} = 2^{1/3} &=& \sqrt{\dfrac{a'^2+b'^2 + c'^2}{a^2+b^2 + c^2}} \\
\Leftrightarrow 2^{2/3} &=& \dfrac{\beta^2+2}{3}\, .
\end{aligned}
\end{equation}
Solving, we find $\beta \sim 1.66$, and thus $c/a = 1/\beta \approx 0.6$.

This predicts an axis ratio $c/a$ slightly smaller than the one found; on the other hand, in the preceding derivation we assumed that all of the change in size occurred along the major axis, whereas the previous section showed that all axes grow to some degree, not just those in the plane of the merger.

Overall, for the range of orbital parameters we have tested, the axis ratios $c/a$ and $c/b$ scale roughly linearly with $\kappa$ and $\lambda$. This simple result suggests that shape changes are relatively easy to understand, and that the details of the initial density profiles are not important, provided the internal energies of the initial haloes are appropriately accounted for.

%%%%%%%%%%%%%%%%%%%%%%%%%%%%%%%%%%%%%%%%%%%%%%%%%%%%%%%%%%%%%%%%%

\section{Conclusions} \label{sec:Discuss}

We have performed a large number of idealized simulations of mergers between isolated haloes with realistic density profiles, to determine what dictates the structure of the remnant in major halo mergers. In this first paper, we describe our IC generation and convergence tests, and then consider the  size and shape of the final remnant, which we find is reasonably well described by a triaxial ellipsoid with axes $a > b > c$. The shape of the remnant is mainly determined by the orbital parameters of the merger, with the energy and angular momentum of the orbit controlling the axis ratios $c/a$ and $c/b$, respectively. The size of the remnant depends mainly on the energy of the orbit, although there is some dependence on the initial halo profile as well. The overall spin of the remnant is also determined by the orbit, through conservation of angular momentum, though the remnant does not generally rotate as a solid body. The radial separation, $r_{sep}$, and the initial velocity, $v_0$, do not have a direct effect on the size or shape of the final halo remnants.

We can interpret our results most simply in terms of the scaled energy parameter $\kappa$ and the dimensionless spin parameter $\lambda$. The former is the net internal energy available to the remnant, relative to its initial energy, and normalized by the overall scaling factor expected if the mean density is conserved while the mass doubles (cf.~equation~\ref{eq:kapp_def}). The latter follows the usual cosmological definition (equation~\ref{eq:lam}). In terms of these variables, we find that the minor-to-major axis ratio $c/a$ scales roughly linearly with relative energy $\kappa$. Mergers with less (negative) total energy (i.e.~low values of $\kappa$, equivalent to merging from large initial separations) produce more elongated remnants, while mergers from smaller initial separations produce rounder remnants. For the `scale invariant' value $\kappa =1$, mergers produce remnants with $c/a \sim 0.7$, as expected from a simple analysis of the energy available along each axis.

The  minor-to-intermediate axis ratio $c/b$ depends mainly on the angular momentum of the original encounter, scaling roughly linearly with the spin parameter $\lambda$. High spin mergers produce oblate, disky remnants that are almost axially symmetric. As the spin parameter decreases, the remnants become progressively more prolate, eventually becoming non-rotating, elongated objects in the limit of radial encounters.

These results are consistent with the previous study of \cite{moore2004}, which found that a radial merger produced a prolate remnant, while a more circular encounter produced a disky remnant, although we extend these results to a much wider range of ICs. Similarily, \cite{mcmillan2007} found that more radial orbits resulted in more prolate remnants. We find that the shape of the final remnant does not depend on the detailed density profile of the initial halo models; this is somewhat contrary to what was found by \cite{fulton2001} and  \cite{mcmillan2007}, who both suggest that shallow cusps produce prolate remnants, and steep cusps produce oblate remnants. This discrepancy is likely because even at fixed orbital energy, the scaled energy parameter $\kappa$ will be different for different ICs if they have different internal energies. We suspect that comparing their results at the same value of $\kappa$ would show no dependence of shape on the initial profile.

Interestingly, we have found that binary equal-mass mergers between spherical haloes rarely result in remnants with shape ratios less than 0.6, while  cosmological haloes typically have shape ratio $c/a<0.7$ \citep[e.g.][]{jing2002,allgood2006,despali2014,bonamigo2015}. This could be because of the intrinsic initial shape of the density perturbations, or because multiple successive mergers often occur along the same filament. We wish to explore this further by considering mergers between two non-spherical haloes along their major axes. Since generating isolated ICs for non-spherical haloes is not straightforward, we can use the remnants from the binary mergers presented in this work as the new ICs, as proposed by \cite{moore2004}.

It should be emphasized that this work only considers equal-mass mergers, which are relatively rare. The rate of mergers per halo decreases with mass ratio, and 1:1 mergers are expected to occur at a rate of approximately 0.1 mergers per unit redshift \citep{fakhouri2010}; non-equal-mass mergers are thus much more common. It is therefore interesting to discuss briefly how our results are expected to extend to non-equal-mass mergers (this will also be the focus of a future study). The parameters $\lambda$ and $\kappa$ defined in this work can also be calculated for non-equal-mass mergers. Overall, we expect that the qualitative results found here will extend to non-equal-mass major mergers. Size should scale inversely with $\kappa$, $c/a$ should scale with $\kappa$, and $c/b$ should scale with $\lambda$. Additionally, the relations derived for the change in the size $a$ in equation~\ref{eq:a_predict}, and the prediction for $c/a \,(\kappa=1)$ in equation~\ref{eq:beta} can be extended to non-equal-mass mergers. This implies, however, that the exact relationship between $c/a$ and $\kappa$ is dependent on the mass ratio of the merger. In the limit of a very large mass ratio, the larger halo will remain unaffected by the merger, and thus $c/a \approx 1$ when $\kappa = 1$.

Overall, there are several caveats to our conclusions. The first is that our ICs represent a great simplification of the typical cosmological situation. In a cosmological setting, haloes are almost never completely isolated, and major mergers between single pairs of haloes are rare. After initial experiments analyzing realistic mergers directly in their cosmological context, we reduced our study to the simplest possible configuration, finding that even simple mergers are complex enough to warrant separate treatment. In future work, we will consider how these results extend to more complicated merger situations such as multiple mergers, or smooth but anisotropic accretion, with the goal of understanding fully the dependence of halo shape on mass and environment that has been measured in cosmological simulations \citep[e.g.][]{jing2002,allgood2006, maulbetsch2007,despali2014,lee2017,vega2017}.

Further, to apply our results to observations, we must also account for baryonic effects. These have been studied before in isolated mergers \citep{aceves2006,kazantzidis2006}. These authors found that the shape of the final merger remnant within the virial radius was similar, whether the merger was simulated using dark matter only, or in full hydrodynamic simulations including baryons. More generally, some hydrodynamic simulations find that baryons make haloes rounder at small radii \citep[e.g.][]{butsky2016}, while others find that they have less effect, at least on cluster scales \cite[e.g.][]{sereno2018}, so further work on this subject is needed. 

Since observations are beginning to place constraints on the shapes of individual galaxy clusters, this is an obvious area in which to pursue the development of next-generation cosmological tests based on structural properties. It would also be interesting to split cluster samples by projected or 3D shape, and compare their mean galaxy content and substructure, to establish a connection between final states and past merger history, for large numbers of systems. Shape and internal structure may also be relevant in understanding the X-ray scaling relations, using the offset from the mean relations versus shape as a probe of how cluster thermodynamics evolve after a major merger \citep[e.g.][]{poole2007}. Finally, in the longer term, the structural properties of individual haloes may be used to probe the statistics of the surrounding density field, including both the spatial anisotropy of the region around the local density peak, and the angular momentum distribution of this region. These properties of the density field should in turn be sensitive to non-Gaussianity and other more subtle aspects of structure formation.

The results presented in this paper should form the basis for a full model of how a halo's shape changes as it grows through mergers and accretion. Such a model may in turn allow semi-analytic predictions of the full distribution of halo shapes as a function of cosmological parameters. In the shorter term, we will use the ICs and analysis tools established here to study the evolution of the density profile and the concentration parameter in major halo mergers. This will be the subject of the second paper in the series \citep{drakos2018}.

\section*{Acknowledgements}
NED acknowledges support from NSERC Canada, through a postgraduate scholarship. JET acknowledges financial support from NSERC Canada, through a Discovery Grant. The authors also wish to thank the Michael Hudson, Julio Navarro, and the anonymous referee for useful comments.

\clearpage
\bibliographystyle{mnras}
\bibliography{MajorMerger2}
% Don't change these lines
\bsp	% typesetting comment
\label{lastpage}
\end{document}